%% file: main.tex
\documentclass[10pt,twocolumn,twoside]{IEEEtran}

\usepackage[dvipsnames]{xcolor}
\usepackage{comment}
\usepackage{tikz}
\usetikzlibrary{external}
\usetikzlibrary{mindmap,trees}
\usetikzlibrary{arrows}
\usetikzlibrary{calc}
\usepackage{subcaption}
\usepackage[ruled,linesnumbered]{algorithm2e}
\graphicspath{{./Figures/}}
\usepackage[font=small,labelfont=bf]{caption}
\usepackage{float}

\usepackage{amsthm}

\usepackage{amsmath,amsfonts,amssymb,color}

\usepackage{epsfig}
\usepackage{psfrag}
\usepackage{array}
\usepackage{physics}
\usepackage{epstopdf}
\usepackage{tikz, cite}
\usepackage[export]{adjustbox}
\usepackage{graphicx} 
\usepackage{bmpsize}

\allowdisplaybreaks

\newif\ifdraft
\drafttrue 

\newcommand{\mustafa}[1]{\ifdraft{\textcolor{red}{MA: #1}}\fi}

\newcolumntype{L}{>{\centering\arraybackslash}m{5cm}}
\newcolumntype{c}{>{\centering\arraybackslash}m{3cm}}

\usepackage{pifont}
\newcommand{\xmark}{\ding{55}}
\usepackage{hyperref}
\usepackage[none]{hyphenat}
\hypersetup{
     linkcolor    = red,
     colorlinks   = true,
     citecolor    = blue
}
\usepackage{cleveref}
\usepackage{tipa}
\usepackage{algorithmic}

\newcommand{\Section}[1]{\vspace{-8pt}\section{#1}\vspace{-4pt}}
\newcommand{\Subsection}[1]{\vspace{-8pt}\subsection{#1}\vspace{-2pt}}

\newcommand{\name}{{\sc IBDASH}\xspace}
\newcommand{\nameexpansion}{\textbf{I}nterference \textbf{B}ased \textbf{D}AG \textbf{A}pplication \textbf{S}c\textbf{H}eduler\xspace}
\newcommand{\namepronunciation}{\textipa{["\={\i}b-dash]}}

\begin{document}
\sloppy

\pagestyle{plain}
\title{DAG-based Task Orchestration for Edge Computing}

\author{Xiang Li$^\alpha$, Mustafa Abdallah$^\alpha$, Shikhar Suryavansh$^\psi$,  Mung Chiang$^\alpha$, Saurabh Bagchi$^\alpha$ \\
$\alpha$: Purdue University; $\psi$: Cisco Systems
}

\maketitle

\begin{abstract}
As we increase the number of personal computing devices that we carry (mobile devices, tablets, e-readers, and laptops) and these come equipped with increasing resources, there is a vast potential computation power that can be utilized from those devices. Edge computing promises to exploit these underlying computation resources closer to users to help run latency-sensitive applications such as augmented reality and video analytics. However, one key missing piece has been how to incorporate personally owned unmanaged devices into a usable edge computing system. The primary challenges arise due to the heterogeneity, lack of interference management, and unpredictable availability of such devices. In this paper we propose an orchestration framework \name, which orchestrates application tasks on an edge system that comprises a mix of commercial and personal edge devices. \name targets reducing both end-to-end latency of execution and probability of failure for applications that have dependency among tasks, captured by directed acyclic graphs (DAGs). \name takes memory constraints of each edge device and network bandwidth into consideration. 
To assess the effectiveness of \name, we run real application tasks on real edge devices with widely varying capabilities. We feed these measurements into a simulator that runs \name at scale. Compared to three state-of-the-art edge orchestration schemes, LAVEA, Petrel, and LaTS, and two intuitive baselines, \name reduces the end-to-end latency and probability of failure, by 14\% and 41\% on average respectively. The main takeaway from our work is that it is feasible to combine personal and commercial devices into a usable edge computing platform, one that delivers low latency and predictable and high availability. 

\end{abstract}

\IEEEpeerreviewmaketitle
\input{Introduction}

\input{Problem_Statement}
\input{Pre_Not}

\input{Framework}
\input{Evaluation}

\input{Relatedwork}
\input{Discussion}
\input{Conclusion}

\bibliographystyle{IEEEtran}
\bibliography{refs}

\end{document}

%% file: Introduction.tex
\section{Introduction}

There has been a surge of latency-sensitive applications running on user-generated streaming data, such as augmented reality and video analytics. Such a surge has driven the wide popularity of edge computing since it offers low latency by performing the computation near the source of the data and offers scalability by distributing the workload among multiple edge devices. 
After some notable innovations in academic publications over the last few years, we have started seeing the growth of small, edge-located data centers managed by infrastructure providers such as Amazon~\cite{lambda_edge}, Microsoft~\cite{ms_edge} and Google~\cite{google_edge}.
We refer to such devices as ``\textbf{Commercial Edge Devices (CEDs)}".
Being commercially managed, CEDs are expected to be available over extended periods and achieve reasonably low latencies. However, they have the drawbacks of incurring \$ cost and still not being widely available. Almost all existing literature on edge computing systems implicitly deals with CEDs as they assume the above desirable properties.

Edge computing systems could also comprise personal devices such as laptops, tablets, and mobile phones. This trend is increasing as such devices are becoming more ubiquitous and are increasing in their compute power. By the end of 2020, it is estimated 6.06B smartphones were in use globally, which is three times the number of PCs, and it is expected to keep growing at a 4\% rate and hit 7.69B by 2026~\cite{personal_edge}. Moreover, smart devices now have unseen storage and processing power and this trend is continuing. For example, a 2010 Samsung Galaxy S only had just 512MB of RAM and 8GB storage with a single core at 1 GHz, but the 2021 Galaxy Z Fold3 comes with 12GB of RAM and 256GB storage with 8 cores clocking at a maximum of 2.84 GHz. 
We call such devices that may be pulled into an edge computing system as ``\textbf{Personal Edge Devices (PEDs)}".
Running latency-sensitive applications on PEDs
is appealing as they are often closer to the user than CEDs, with the same user often carrying multiple PEDs. Furthermore, with the right kind of incentive schemes, the usage of PEDs can be at zero cost.
On the other hand, such devices are expected to have sporadic availability and have little to no management of contention that can arise due to multiple co-located applications. Therefore, one has to carefully manage such PEDs in an edge computing system to achieve reliable and low latency executions.

In this paper, we present the design of an edge computing task scheduling scheme that we call \textbf{\name}\footnote{\name stands for \nameexpansion and is pronounced as \namepronunciation.} that combines PEDs and CEDs into one system to leverage the benefits of both types of devices. In particular, \name introduces a method to schedule complex latency-sensitive applications, whose tasks (stages) have dependencies and can be represented by directed acyclic graphs (DAGs).
The scheduling happens among available PEDs and CEDs to reduce the end-to-end execution latency of each application and the probability of application failure, while accounting for the dependencies among different stages of the application. For example, a video analytic application may do scene change detection and pass onto a second task that does object detection only if there is a scene change. \name also takes into account the interference on a particular device from multiple co-located applications. This is particularly important because in our target class of devices, there do not exist good hardware mechanisms for avoiding contention, as compared to server-class devices~\cite{serverclass}. 

Table \ref{tbl:related_work} compares the features of our proposed solution \name with prior related works. In particular, prior works such as 
LAVEA~\cite{lavea} and Petrel~\cite{petrel} propose their orchestration schemes that target client-edge offloading to provide low-latency video analytic and randomized load balancing by leveraging the ``power of two" choice~\cite{power_o2} to randomly choose two edge devices and allocate the task to the one which gives better performance. Moreover, LaTS~\cite{lats} allocates the task to the device that has the shortest estimated latency based on a pre-profiled latency-CPU usage model. On the other hand, JCAB~\cite{jcab} effectively balances the accuracy and energy consumption while keeping low system latency by jointly optimizing configuration adaption and bandwidth allocation.~\cite{rtod} proposes a system that employs low latency offloading techniques jointly with pipeline decoupling methods and fast object tracking methods to enable accurate object detection. However, among these frameworks, some do not consider the heterogeneity of edge devices at all~\cite{jcab, rtod} and among those that consider the heterogeneity of edge devices,~\cite{lavea} considers the heterogeneity of CPU architectures,~\cite{lats} considers the heterogeneity of CPU and GPU mix, and~\cite{petrel} considers the mix of devices as a cloudlet entity. {\em None of these works considers the mix of PEDs and CEDS
.} Moreover, most of them (except LaTS~\cite{lats}) do not address the interference of co-located tasks on the same edge device. The previous work IBOT~\cite{suryavansh2020bot} 
proposed an orchestration scheme that takes the interference among tasks on each edge device into account and orchestrates an optimal execution strategy that jointly optimizes both execution latency and probability of failure. However, IBOT treats each task separately and does not take into account dependencies among tasks within an application and it also fails to address the memory constraints of each edge device since some tasks may require loading models into memory to successfully carry out the task execution.

\begin{table}[t]
\vspace{0.3mm}
\caption {A comparison between the prior related works and our system in terms of the available features. 
\name offers DAG support, failure reduction and memory consideration. 
}
\label{tbl:related_work}
\centering
\resizebox{\columnwidth}{!}
{%
\begin{tabular}{|l|l|l|l|l|l|l|l|}
\hline
\multicolumn{1}{|l|}{\text{\bf Framework}}
& \multicolumn{1}{l|}{\bf \shortstack{Failure\\ reduction}}
& \multicolumn{1}{l|}{\bf \shortstack{Supporting \\ DAG }}
& \multicolumn{1}{l|}{\bf \shortstack{Heterogeneous\\ edge devices}}
& \multicolumn{1}{l|}{\bf \shortstack{Memory \\ Consideration }}
& \multicolumn{1}{l|}{\bf \shortstack{Low \\ latency }}
& \multicolumn{1}{l|}{\bf \shortstack{Orchestration\\ overhead \\ reduction}}\\
\cline{1-7}
\hline
LaTS~\cite{lats} & \xmark & \checkmark & \checkmark & \xmark & \checkmark & \xmark\\
\hline
Petrel~\cite{petrel}  & \xmark  & \xmark  &  \checkmark & \xmark & \checkmark & \checkmark \\
\hline
LAVEA~\cite{lavea} & \xmark & \checkmark & \checkmark & \xmark & \checkmark & \xmark \\
\hline
Edge Object Detection ~\cite{rtod} & \xmark & \xmark & \xmark & \xmark & \checkmark & \xmark \\  
\hline
JCAB~\cite{jcab} & \xmark & \xmark & \xmark & \xmark & \checkmark & \xmark \\  
\hline
IBOT~\cite{suryavansh2020bot} & \checkmark & \xmark & \checkmark & \xmark & \checkmark & \checkmark \\  
\hline
\name (This work) & \checkmark & \checkmark & \checkmark & \checkmark & \checkmark & \xmark \\  
\hline
\end{tabular}}
\vspace{-0.1in}
\end{table}

\begin{figure}[t]
\begin{center}
\includegraphics[scale=0.32]{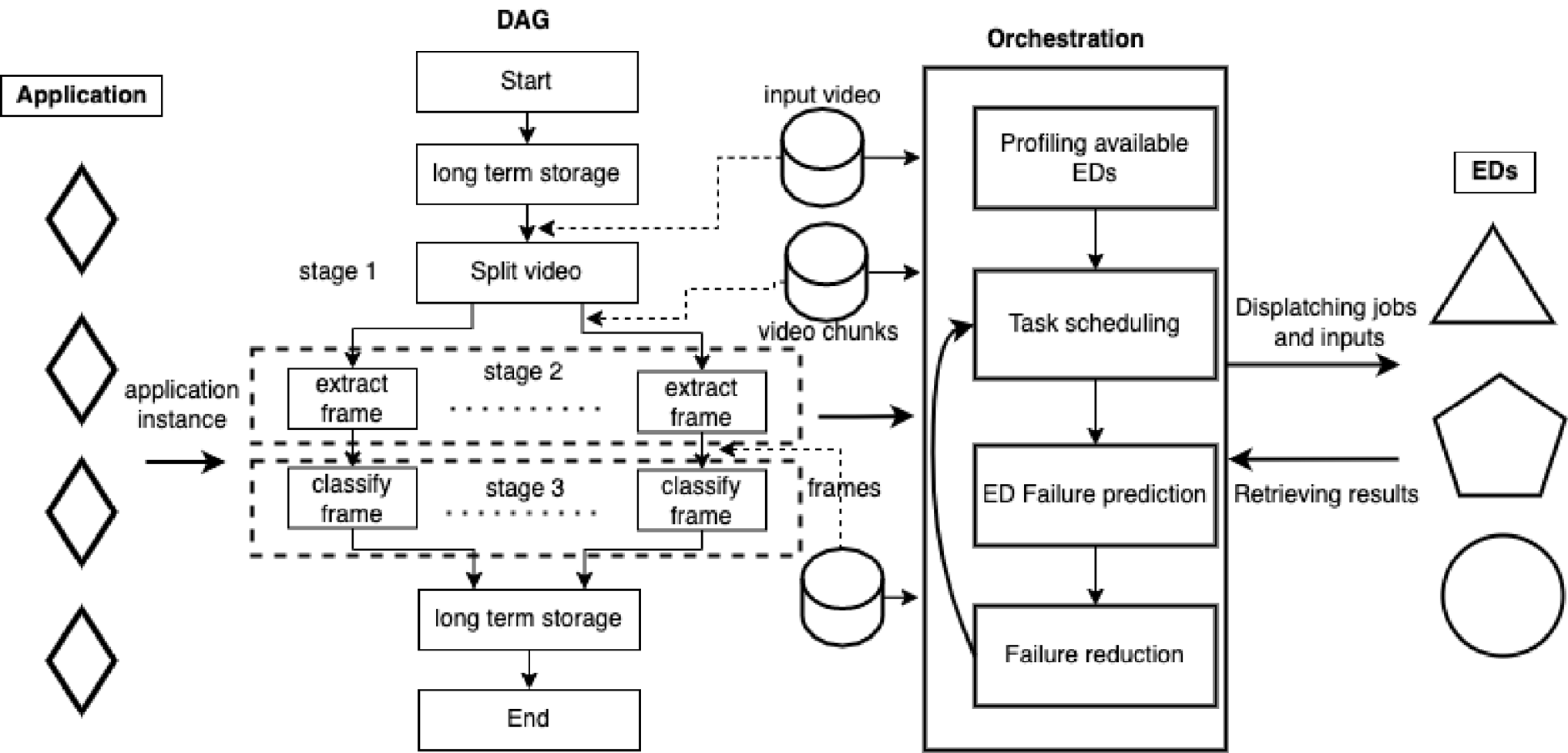}
\end{center}
\caption{A system overview of \name orchestration scheme for DAG based application. All edge devices in the network need to be profiled before task allocation. \name  do pre-processing the tasks in the application and puts them into different stages, then the scheduler orchestrates the allocation for each task's execution to reduce execution latency and probability of failure.}
\label{fig:sys_overview}
\vspace{-3mm}
\end{figure}
Figure \ref{fig:sys_overview} shows an overview of our proposed solution \name.
Each application instance from the end-user consists of one or more tasks, which may have dependencies among them. 
For example, in Figure \ref{fig:sys_overview}, we show a DAG example of a video analytics application that shows control and data dependencies among stages. The application is explained in detail in Section\ref{sec:testing_app}.

In our evaluation, we compare \name  with two intuitive baselines, Random allocation and Round Robin, 
and three state-of-the-art solutions, LAVEA~\cite{lavea}, Petrel~\cite{petrel}, and LaTS~\cite{lats}. To test our framework, we use four applications that span various DAG structures from different application domains. Some tasks require models for their execution and some do not. 
For example, if the task performs object recognition, a pre-trained model is  needed on the designated edge device before the task can start running. 
Compared with existing schemes, \name reduces the average service time of applications
by 
14\% compared to the best baseline scheme. Concurrently with reducing service time, \name reduces the average probability of failure for the application by 
41\% compared to the best baseline scheme.

In summary, this paper makes the following contributions:
\begin{enumerate}
    \item We propose an orchestration framework \name, an interference-based dynamic task orchestration scheme that executes DAG-based user applications in a heterogeneous edge computing environment with low latency and low probability of failure.
    \item To increase the reliability of edge computing, our solution proposes adding redundancy into the platform by replicating the tasks allocated to devices with a high probability of failure to multiple edge devices.
    \item We propose a device availability prediction model and validate it through the data collected by~\cite{mobility}.
    \item We validate our model via extensive simulation of application instances that arrive randomly within a time period
    and shows its superiority in reducing average application service time and probability of failure.
\end{enumerate}

%% file: Problem_Statement.tex
\section{Problem Statement}\label{sec: ps}
The combination of CEDs and PEDs, task dependencies within applications, and sporadic availability of PEDs pose unique challenges that have not been addressed in the edge computing literature. We discuss the four primary challenges, whose solutions bring out the novelty in \name. 

\noindent \textbf{Substantial heterogeneity in computational capacity:} PEDs such as laptops, tablets and mobile devices can have a substantial variation in their compute power, memory, etc. For example, in the current smartphone market, Samsung, Apple, and Xiaomi contribute 20\%, 14\%, and 13\% respectively to the market share and others contribute the remaining 53\%. Within each brand, there  is a  wide range of devices with different capabilities that target different customers.
The penetration of different brands in different markets and economies varies widely leading to a natural heterogeneity in the PEDs. 

\begin{figure}[h]
\centering
\includegraphics[scale=0.31]{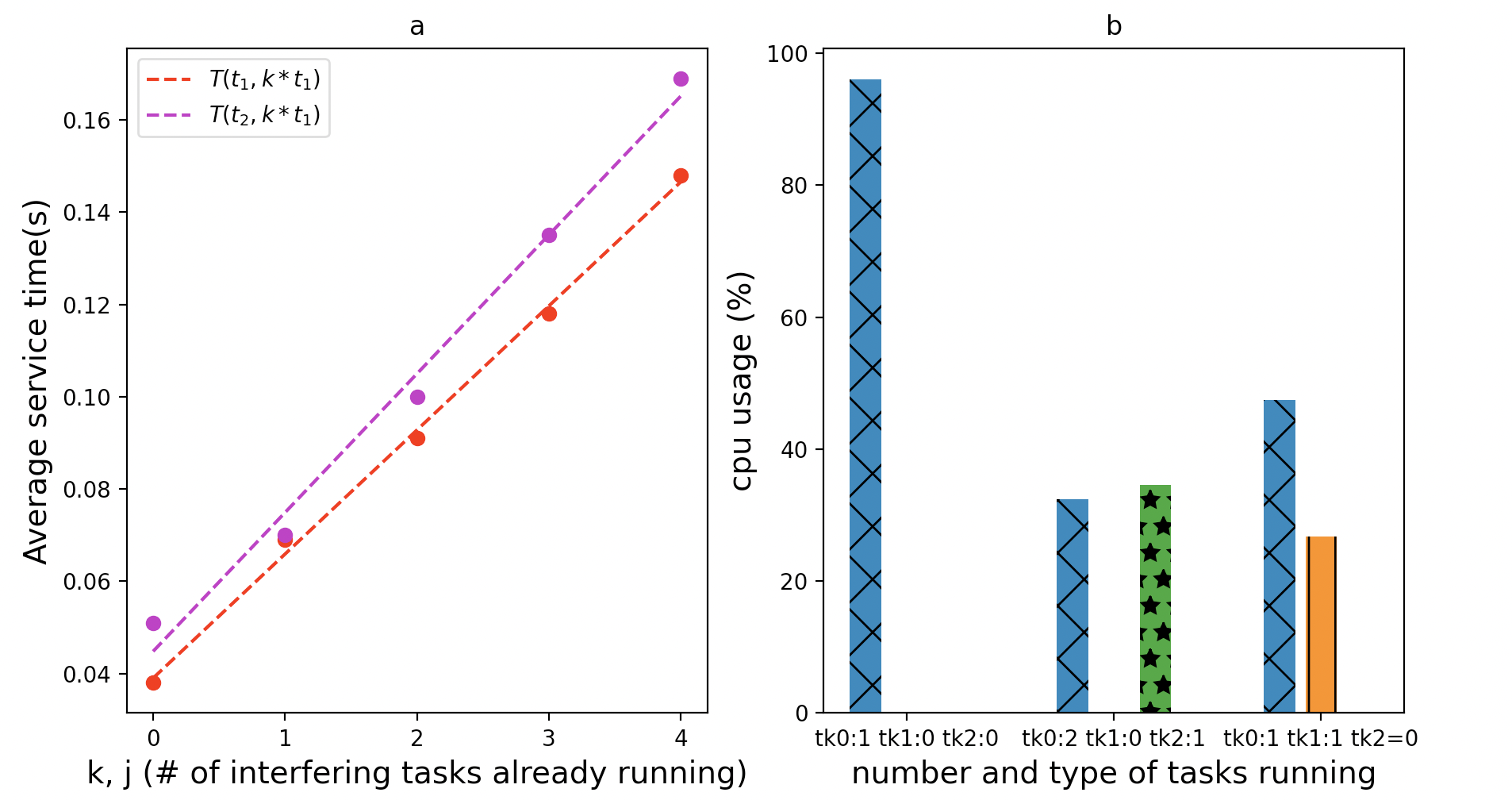}
\caption{\textbf{(a) The heterogeneity in interference among different tasks. (b) The  CPU usage of each task varies when different tasks are running in the background.}}
\label{fig:tk_inter}
\end{figure}

\noindent \textbf{Heterogeneity in task interference pattern:} Different tasks, when running on the same edge device, interfere with each other affecting their execution time. There is heterogeneity in the interference experienced by different tasks on an edge device. For example, suppose that we have three tasks where the first task ($t_0$) loads a set of images, the second task ($t_1$) performs convolution on the pre-loaded images, and the third task ($t_2$) rotates the processed images. For such a scenario, Figure \ref{fig:tk_inter} shows the different interference patterns for different task types and the different CPU usages for different task types. In Figure \ref{fig:tk_inter}a, we see that the interference pattern for $t_{1} - t_{1}$ is different from the interference pattern for $t_{2} - t_{1}$. Figure \ref{fig:tk_inter}b shows that the CPU usage for each task under three different scenarios also varies. 
The interference patterns and CPU usage differences can be due to multiple reasons, such as resource contention~\cite{resource_contention}, priority scheduling~\cite{priority_scheduling}, etc. The main insight to note is that the different interference patterns among tasks result in different service times.

 \noindent \textbf{Sporadic availability of PEDs:} Due to the unmanaged nature of the PEDs, their availability in the network can be hard to predict. For example, in a more predictable scenario such as a classroom setting, when students leave the classroom, their laptops and mobile devices will not be available anymore so any tasks that are scheduled close to the end of class will experience a high probability of failure. However, in other scenarios where people come and go less predictably 
(e.g., a university library), it would be hard to predict the availability. For instance, in~\cite{mobility} we did an experiment to track the availability of the mobile devices of students on university campus.
The results show that the probability of failure for mobile devices (mobile devices disconnect from the crowdsensing framework) increases with the length of time that elapses since they connected to the framework.

\noindent \textbf{DAG-based application orchestrations:} The dependency among tasks within the same application adds one more layer of complexity into the orchestration problem as the execution of tasks need to follow a certain order. Some can be executed in parallel and some cannot.  
Prior work~\cite{edge_overview} shows that several partition algorithms~\cite{clonecloud,maui} are developed to achieve different optimization goals such as saving energy~\cite{save_energy,deep_decision}, reducing execution latency~\cite{deep_decision}. Our proposed framework \name utilizes the structure within DAG-based application where it explicitly characterizes the task flow and data flow. Then, we orchestrate the task allocation to reduce the overall end-to-end latency and probability of failure. The significant prior work on DAG scheduling in the cloud is less relevant in our context as our EDs are more heterogeneous and less predictable in their availability.

%% file: Pre_Not.tex
\section{Preliminaries and Notations}\label{sec: notations}



\subsection{Feasibility of PEDs: A survey}
\begin{figure*}[ht]
        \centering
        \begin{subfigure}[h]{0.66\columnwidth}  
            \centering 
            \includegraphics[width=\columnwidth]{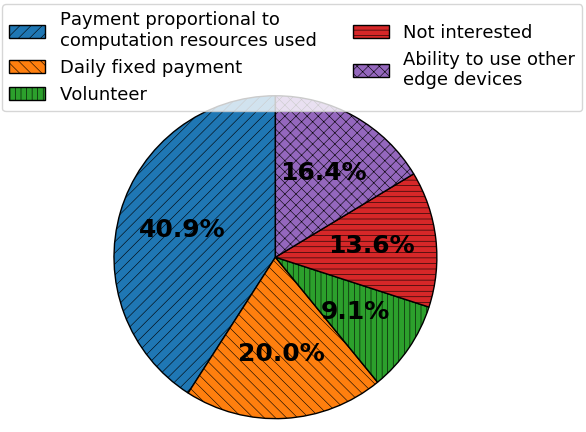}
            \caption[]{Preferred incentive model}
            \label{fig:incentive_model}
        \end{subfigure}
        \hfill
        \begin{subfigure}[h]{0.66\columnwidth}  
            \centering 
            \includegraphics[width=\columnwidth]{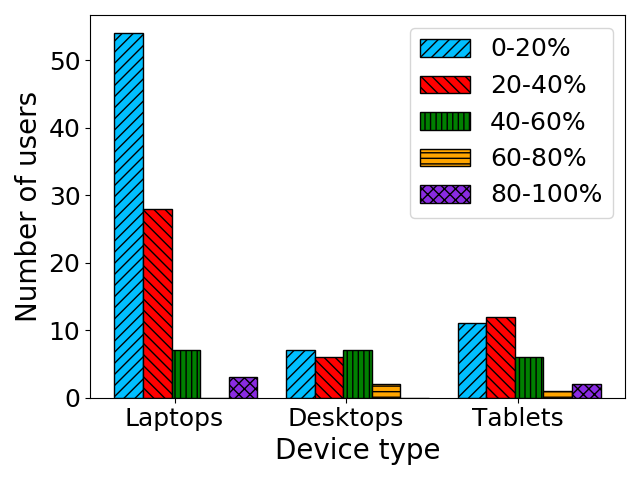}
            \caption[]{Percentage of CPU resources willing to share}
            \label{fig:resource_shared}
        \end{subfigure}
        \hfill
        \begin{subfigure}[h]{0.66\columnwidth}
            \centering
            \includegraphics[width=\columnwidth]{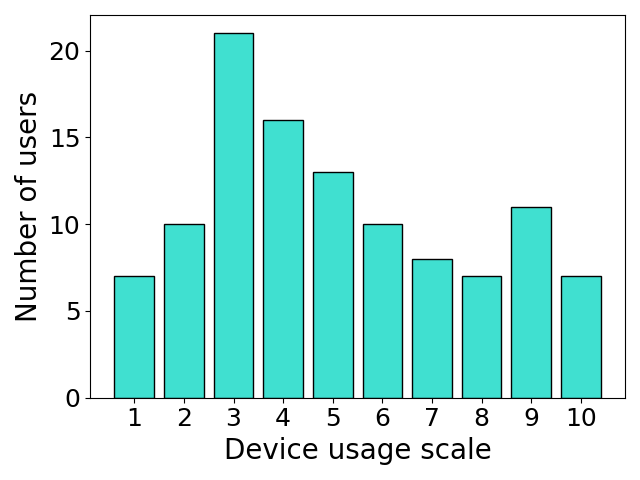}
            \caption[] {Device usage (1: Low usage, 10: Very high usage)}
            \label{fig:device_usage}
        \end{subfigure}
        \vspace{-5 pt}
        \caption[]
        {User survey results ($N$ = 110)} 
        \label{fig:survey_results}
        \vspace{-5 pt}
    \end{figure*}

We conducted a user study with 110 participants from USA and India that are engaged in diverse fields such as educators, software professionals, students, engineering professionals, etc. to understand their willingness to share their computing devices (e.g., laptops, desktops, tablets, etc.) as edge devices. Figure \ref{fig:incentive_model} shows that 86.4\% of the participants are willing to share their devices under one of four proposed incentive schemes. Only 13.6\% of the participants were not interested in sharing their devices at all, primarily due to privacy and security concerns. Moreover, Figure \ref{fig:resource_shared} indicates that the majority of the participants are willing to share 0-40\% of their CPU resources. The amount of CPU resources people are willing to share varies as well depending on the device type. As shown in Figure \ref{fig:device_usage}, we obtained a double Gaussian device usage pattern with peaks at 90\% and 30\% of usage indicating that most people either use their devices very heavily (video editing, running sophisticated software, etc.) or use them only for computationally light applications such as browsing, reading, etc. The average usage across all users was 50.9\%, thereby supporting our claim that a lot of devices are not utilized to their capacity.

Now, we introduce the notations and terms used in our framework \name and show them in Table \ref{tbl:notation}. In particular, we explain the DAG, edge computing, task orchestration, and main system metrics.

\begin{table}[h]
\caption {The list of symbols and definitions used in our work.}
\label{tbl:notation}
\centering

\begin{tabular}{|l|l|}
\hline
\multicolumn{1}{|c|}{\text{\bf Symbol}}
& \multicolumn{1}{c|}{\bf Definition} \\
\cline{1-2}
\hline
$T = \{T_{1}, T_{2}, \hdots, T_{N}\}$ & Types of tasks for a given application\\
\hline
$ED =\{ED_{1}, \hdots, ED_{N}\}$ & Total number of edge devices\\
\hline
$S =\{S_{1}, \hdots, S_{N}\}$ & Number of stages in DAG\\
\hline
$G=(V_{i},E_{j}),\ V_{i} \in T$  & DAG representation of the application\\ 
\hline
$L(T_{i})_{ED_{j}}$ & Execution latency of $T_{i}$ on ED j \\
\hline
$L(M(T_{i}))_{ED_{j}}$ & Model upload latency for $T_{i}$ to ED j\\
\hline
$H(ED_{j})$ & Memory available on edge device j\\
\hline
$H(T_{i})$ &  Memory required for $T_{i}$\\
\hline
$L(T_{i})_{d}$ & Data transfer latency for Task i input\\
\hline
$L(T_{i})$ & End-to-end latency for task $T_{i}$\\
\hline
$L(G)$ & End to end latency for the application\\ 
\hline
$T(i)_{d}$ & Input data for task $T_{i}$\\
\hline
$L(S_{i})$ & End-to-end latency of stage $i$\\
\hline
$P(T_{i})$ & Placement of task $T_{i}$\\
\hline
$D(T_{i})$ & Dependency of task $T_{i}$ \\
& in terms of other tasks\\
\hline
$P(G)$ & Placement of each task in graph G\\
\hline
$P_{f}(G)$ & Probability of failure of application,\\
& given by graph G\\
\hline
$T_{rep}$ & Tracker for the number of replications\\
\hline
$\beta$ & Probability of failure threshold\\
\hline
$\gamma$ & Threshold on the Replication degree\\
\hline
$F(T_{i})$ & Probability that $T_{i}$ fail\\
\hline
$B$ & Network bandwidth\\
\hline
$M(T_{i})$ & Model required for $T_{i}$\\
\hline
$Ed_{info}$ &  Total and free space on each ED\\ 
\hline
$M_{info}$ & Available models on each edge device\\
\hline
$Task_{info}$ & \# executing tasks and 
 types on each ED  \\
\hline
$WeightS$ & Weight score of joint optimization \\
\hline
$WeightSnew$ & Weighted score after PF reduction\\
\hline
\end{tabular}
\vspace{-5pt}
\end{table}

\Subsection{Directed Acyclic Graphs}
Each application is represented as a directed acyclic graph (DAG) $G = (V,E)$. The set of nodes $V$ represents the individual tasks in an application instance, while the set of edges $E$ characterizes the dependency among those tasks. The dependency can mean both execution order dependency or data dependency. In particular, an edge from task $v_{i}$ to task $v_{j}$ indicates that $v_{i}$ needs to be finished before the start of $v_{j}$.

\Subsection{End-to-end Latency and Probability of Failure}
Throughout the paper, we used the terms end-to-end latency and probability of failure to describe the goal of joint optimization. End-to-end latency is defined as the time from when the first task in the DAG-based application starts executing till the last task finishes its execution. For this purpose, we assume that the clocks on all devices are synchronized, which can be achieved through means such as~\cite{koo2009tale}. In this paper, we use the average end-to-end latency of hundreds of application instances to evaluate the performance of our orchestration strategy. The term probability of failure is defined as the probability that the application instance did not successfully finish its execution. An application instance may not complete its execution because one of the edge devices becomes unavailable in the middle of task execution, or the owner of the edge device decided to perform some heavy-duty task, which results in the edge device being less responsive.
  
Having explained the problem statement and the main notations used in our work, we now turn our attention to the design of \name.

%% file: Framework.tex
\section{Our Proposed Solution: \name}\label{sec: framework}
To target the problems we listed in Section \ref{sec: ps}, we propose the framework \name, which is an interference-based orchestration scheme for DAG-based applications that aims at jointly optimizing the end-to-end latency and probability of failure for application instances. The rest of this section covers the task interference in our framework, the components of the framework, and our proposed orchestration algorithm.

\Subsection{Interference service time plots}
We define interference as a linear service time plot $T_{i}=m_{j}*k+c_{j}$ that characterizes the execution time of a new task of type $T_{i}$ on $ED_{p}$, given that $k$ tasks of type $T_{j}$ are already running on that edge device. For example, on a given edge device, for a new task $T_{i}$, we can plot $N$ interference plots for every other type of task including $T_{i}$ itself. Therefore, there are overall $N^{2}$ such plots and $N^{2}$ pairs of \textit{m} and \textit{c} values to characterize all interference plots for that edge device. The expected service time of the new incoming task $T_{i}$ on $ED_{p}$, which has $\alpha_{1}$, $\alpha_{2}$, $\hdots$, $\alpha_{N}$ running tasks is given by:

\begin{align}
& f_{i,(1,2,\hdots,N)}(T_{i},(\alpha_{1}T_{1},\hdots,\alpha_{i}T_{i},\hdots, \alpha_{N}T_{N})) \nonumber \\
& = \sum_{j=1}^N f_{ij}(T_i, \alpha_jT_j)
\label{equ:inter}
\end{align}

The above equation assumes that the interference patterns are independent and additive, which we verify experimentally (Section \ref{sec: experiment} Figure \ref{fig:pairwise}). Lower interference coefficient ($m$, $c$ values) of an application for a device means the shorter estimated execution latency for running that application on the device.
A pairwise interference coefficient matrix has been defined as \textit{$ED_{mc}$}, in which each row contains $N^{2}$ pairs of \textit{m}, \textit{c} values for that particular edge device. The element $<m_{ij}, c_{ij}>_{p}$ means that if we want to schedule a new task of type $T_{i}$ while \textit{k} instances of $T_{j}$ are already running on edge device \textit{p}, the estimated service time for $T_{i}$ can be calculated as $k*m_{ij}+c_{ij}$.
We use the matrix \textit{$Task_{info}$} to record the allocation of each task and the estimated time it will be on that edge device, then we can calculate the number of running tasks on each device at a certain time by a simple summation.

\Subsection{Design Components}
The proposed framework contains three main functions, DAG transformation, minimum service time scheduling, and failure likelihood reduction. When a new application instance arrives, the framework first transforms the application's DAG and divides the execution into stages. The advantage of dividing the DAG into stages is that the dependency of the tasks is embedded within the stages and all tasks within the same stage can be executed in parallel. Figure \ref{fig:sys_overview} shows an example of staged DAG for video analytics applications. 
Such staging process is performed through modified Breadth-First Search where the stage of a node is the length of the longest path from the start node.

After staging the DAG, the orchestrator uses the profiling data (saved in matrix $ED_{mc}$) to retrieve the interference coefficients ((\textit{m}, \textit{c}) value pairs) and using $Task_{info}$ matrix to retrieve the number of running tasks of each task type on the edge device of interest. 
The execution latency of the task is estimated using Equation \ref{equ:inter} and denoted as $L(T_{i})_{ED_{p}}$. Besides the execution latency, some tasks may require models to successfully execute the task. Therefore, we introduce the term $M(T_{i})$ to denote the models required for task $T_{i}$ (this may be null in cases) and $L(M(T_{i}))_{ED_{p}}$ to express the corresponding model uploading latency, which depends on both the model size and the network bandwidth $B$. Another important latency factor is the data transfer latency of the input data $T(i)_{d}$ for task $T_{i}$ if it is from other edge devices, and this is expressed as $L(T_{i})_{d}$. The device $p$ that gives minimum latency of the task $T_i$, which is the summation of the minimum execution latency, model uploading latency and data transmission latency for  $T_{i}$. The choice of the edge device is one that minimizes this combined latency and is given by
\begin{align}
& \underset{p}{\text{arg min }} L(T_i) \nonumber \\  
& where \ L(T_i) = L(T_{i})_{ED_p}+L(M(T_{i}))_{ED_{p}}+L(T_{i})_{d} \nonumber \\
& s.t. \ B \leq B_{max},\ H(T_{i}) \leq H(ED_{p}) \nonumber, \ ED_{p}\ \in\ ED \\
\end{align}
Here, $B$ is the current network bandwidth, $B_{max}$ is the maximum available network bandwidth, $H(T_{i})$ is the memory required for $T_{i}$'s execution, including memory to store data and model, and $H(ED_{p})$ is the available memory on $ED_{p}$.

Now, let us define $L(S_{i}) = \max_{T_i \in S_i} L(T_{i})$ as the stage $i$ latency. Therefore, the end-to-end latency of the entire application is the sum of the longest latency task in each stage.
\begin{align}
& L(G) = \Sigma_{i=1}^{i=S}L(S_{i})
\end{align}
In the end, due to the sporadic availability of PEDs, \name adds redundancy to replicate tasks that are assigned to edge devices with a high probability of failure to other edges devices. The goal of this redundant replication is to reduce the average probability of failure of the application instance below a certain threshold or to the minimum probability of failure within the replication degree constraints. For each edge device that was chosen to execute task $T_{i}$, we predict the probability of failure of that edge device during the estimated service time of task $T_{i}$ and this is the probability of failure of task $T_{i}$, which is denoted by $F(T_{i})$. If $F(T_{i})$ is above a certain threshold $\beta$, \name replicates $T_{i}$ on another edge device which gives the next optimal minimum service time. We keep repeating the replication until $F(T_{i})$ is reduced below the probability of failure threshold or the number of replications for $T_{i}$ reaches the replication degree $\gamma$. 
Every task within the application instance needs to be successfully executed so that the entire application can be counted as successful. Therefore, our framework seeks to minimize the probability of failure of every single task. We use $1-F(T_{i})$ 
to denote the probability of success for $T_{i}$, then uses the intersection of the probability of success for all tasks to denote the probability of success for the entire application instance. Such intersection notation captures the dependence among tasks (which is captured by conditional probabilities). In our setup, we used the data to estimate such conditional probabilities of task successes.
\begin{align}
P_{f}(G) = 1 - \cap_{i=1}^{N} (1-F(T_{i}))
\end{align}
The final optimization problem is given by 
\begin{align}
 \min  \alpha L(G)+(1-\alpha)P_{f}(G),
\end{align}
where $\alpha$ is the joint optimization parameter that is controlled by the user to give proper weight for end-to-end latency and probability of failure.

\Subsection{Orchestration algorithm description}
The orchestration algorithm is shown in Algorithm 1.
It greedily examines each task on available edge devices, while concurrently considering the allocation of its prerequisite tasks to minimize application latency and likelihood of failure globally. The algorithm outputs the placement choice for each task in the application on an edge device, PED or CED.

Line 4-17 checks each task in each stage against all available edge devices for estimated latency and consider the extra data transmission and modeling uploading latency globally when tasks with dependency are assigned to different devices. Line 6 calculates the expected execution latency of task $T_{i}$ on $ED_{p}$. Line 7-10 estimates the model uploading latency based on network bandwidth and whether the task requires models for its execution and the model's availability on the targeted edge device. 
 Line 11-14 estimates the input data transmission latency for task $T_{i}$ based on the dependency of task $T_{i}$, which is denoted as $D(T_{i})$. Line 15-16 sums up execution latency, model uploading latency, and data transmission latency then save the end-to-end latency for task $T_{i}$ and the corresponding edge device identifier to a priority queue. By the end of the execution of the for loop from lines 5-17, we will end up with the expected latency of $T_{i}$ on each edge device. Line 18 dequeues from the priority queue and the dequeued item contains the most optimized end-to-end latency placement for task $T_{i}$. Line 19-27 removes not frequently used model to free space and uploads the required models to the targeted edge device and updates the $M_{info}$ structure which is used to keep track of the model availability on each device and $ED_{info}$ structure, which is used to keep track of the memory available on each device. Line 28 calculates the probability of failure of the task on the device based on its dependency then a weighted optimization score is calculated in line 29.
 
 Now, the algorithm checks the probability of failure against the preset threshold $\beta$. If the probability of failure is greater than the threshold and the number of replications for the task is less than replication degree $\gamma$, line 31 dequeues the next item and lines 32-40 recalculate the weighted optimization score with the new probability of failure and the execution latency. If the new weighted optimization score is less than the previous weighted score, we replicate the task. Then repeating this process until the probability of failure is below the threshold $\beta$, or replication degree reached, or the queue is exhausted. Line 42 records the task allocation and line 44 records the longest latency in the current stage and makes sure that tasks from the next stage will not start until the previous stage is finished due to dependency. Finally, line 46 keeps track of the end-to-end latency of the entire application instance.

\begin{algorithm}
    \SetAlgoLined
    \caption{Orchestration Algorithm}
    \SetKwInOut{KwIn}{Input}
    \SetKwInOut{KwOut}{Output}
    \SetKwInOut{KwInit}{Initialization}
    \SetKwInOut{KwRe}{return}
    \KwIn{DAG representing application instance G}
    \KwOut{Task placement $P(T_{i}) \forall T_{i} \in {T_{1},T_{2},\cdots, T_{N}}$, application end-to-end latency $L(G)$}
    \KwInit{$ED_{mc}$, $ED_{info}$, $Task_{info}$}
    $P.init()$ \# init placement structure\; 
    $S = app\_stage(G)$ \# stagerize the DAG\;
    \For{$S_{i} \in S$}{
        \For{$T_{i} \in S_{i}$}{
            \For{$ED_{p} \in ED$ }{
                $L(T_{i})_{ED_{p}} =
                GetEstimatedTime(T_{i},ED_{p})$\;
                $L(M(T_{i}))_{ED_{p}} = 0$\;
                \If{$M(T_{i})\ not\ on\ {ED_{p}}$}{
                    $L(M(T_{i}))_{ED_{p}}=
                    GetMdUpTime(size(M(T_{i})),B)$\;
                 }
                 $L(T_{i})_{d} = 0$\;
                \If{$T(i)_{d}\ not\ on\ \ ED_{p}$}{
                    $L(T_{i})_{d} = GetDTrTime(size(T(i)_{d}),B,D(T_{i}))$\;
                }
                $L(T_{i})=L(T_{i})_{ED_{p}}+L(M(T_{i}))_{ED_{p}} + L(T_{i})_{d}$\; 
                $PQueue.enqueue([ED_{p},L(T_{i})])$
            }  \;
            $ED_{p},\ L(T_{i}) = PQueue.dequeue()$\;
            \eIf{$M(T_{i})\ not\ on\ {ED_{p}}$}{
                \While{$H(ED_{p})\leq H(T_{i})$}{
                    $M_{info}[ED_{p}].removeEnd()$ 
                }   
                $M_{info}[ED_{p}].add(M(T_{i}))$ \;
                $ED_{info}.Update()$
            }{
                $M_{info}[ED_{p}].moveFront(M(T_{i}))$
            }\;
            $F(T_{i}) = GetPf(T_{i},ED_{p},L(T_{i})_{ED_{p}},D(T_{i}))$\;
            $WeightS=\alpha L(T_{i})+(1-\alpha)F(T_{i})$\;
            \While{$F(T_{i}) \geq \beta$ and $T_{rep}<\gamma$}{
                $ED_{p},\ L(T_{i}) = PQueue.dequeue()$\;
                $F(T_{i}) = GetPf(T_{i},ED_{p},L(T_{i}))$\;
                $WeightSnew=\alpha L(T_{i})+(1-\alpha)F(T_{i})$\;
                \eIf{$WeightSnew \leq WeightS$}{
                    $P(T_{i}).add(ED_{p})$\;
                    $WeightS = WeightSnew$\;
                    $T_{rep}++$\;
                 }{
                 $break$
                 }
            }
            $P(T_{i}) = min(\alpha L(T_{i})+(1-\alpha)F(T_{i}))$ \;
        }
        $L(S_{i})$ = $max_{T_{i} \in S}(L(T_{i}))$
    }
    $L(G)=\sum_{i=1}^{i=S}max(L(S_{i})$)
    
    \KwRe{$P(T_i) \forall T_i \in \{T_1,T_2, \cdots,T_N\}$,\ $L(G)$ }
\label{algo:orchestration}
\end{algorithm}

%% file: Evaluation.tex
\begin{figure}[h]
\includegraphics[scale=0.30]{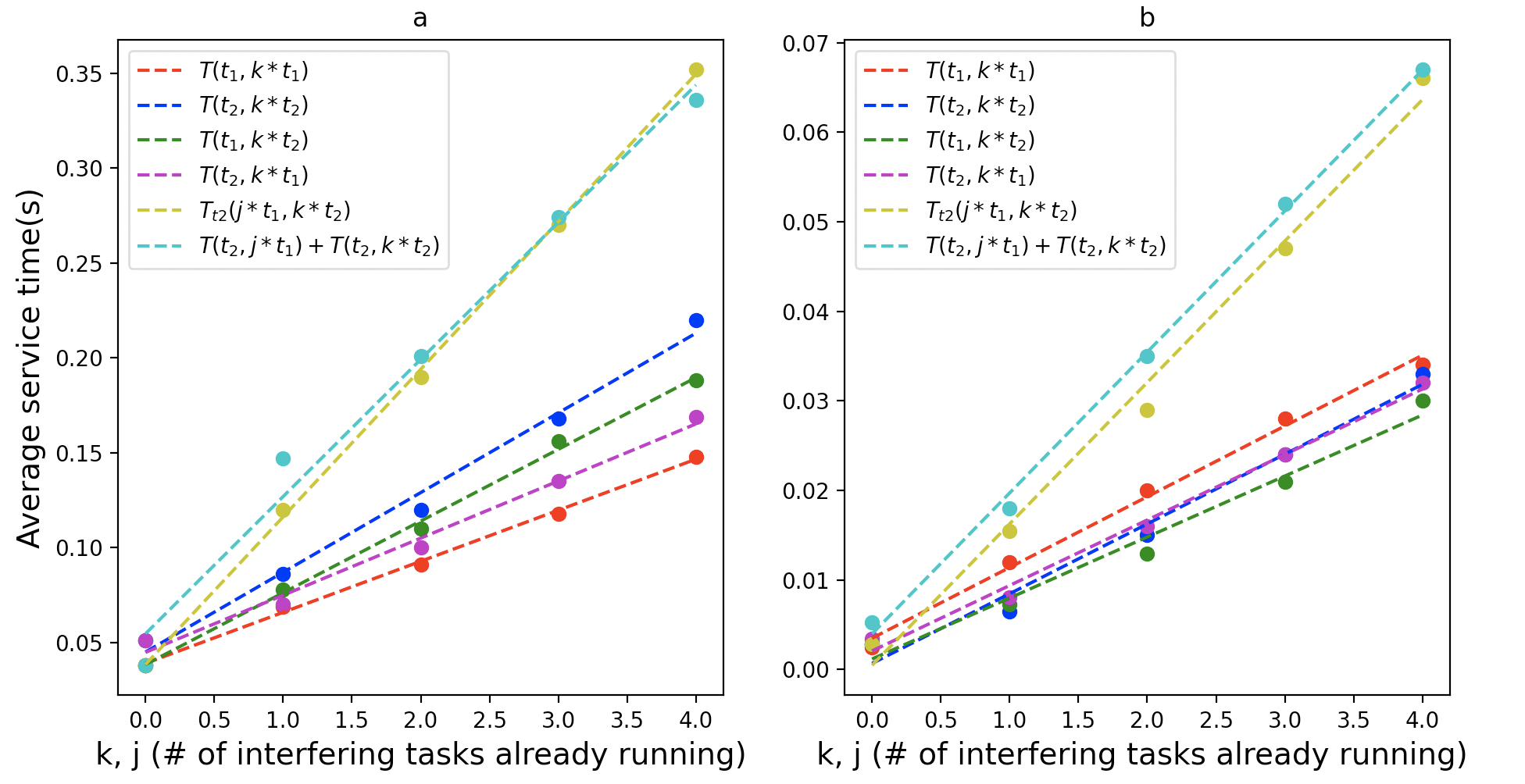}
\centering
\caption{\textbf{Interference pattern verification, $j$,$k$ are the number of tasks of $t_{1}$ and $t_{2}$ running on the ED. (a) Interference pattern on MacBook Pro. (b) Interference pattern on Huawei Nexus 6P.}}
\label{fig:pairwise}
\end{figure}

\Section{Experiments}\label{sec: experiment}
We seek to answer the following evaluation questions:
\begin{itemize}
    \item RQ1: What is the interference pattern among tasks? 
    \item RQ2: How does \name's performance compare to other baseline schemes with respect to end-to-end latency and failure likelihood?
    \item RQ3: How to predict the availability of edge devices?
    \item RQ4: How do the parameters $\alpha, \gamma$ affect the latency and failure likelihood achieved by \name?
\end{itemize}
\Subsection{Interference pattern verification}
In Section \ref{sec: framework}-A, we assume that the interference patterns among tasks are independent and additive. We validate this assumption through an experiment whose result is shown in Figure~\ref{fig:pairwise}. Two computationally intensive tasks have been used to verify our assumption on two different platforms. In particular, task $t_{1}$ represents matrix multiplication of randomly generated floating point entries 100 times 
and task $t_{2}$ denotes inversion of matrices with the above randomly generated floating entries. The two platforms used in this experiment are a Macbook Pro (3.1GHz dual-core Intel core i5 and 8GB  2133 MHz LPDDR3 memory) 
and a Huawei Nexus 6P (Qualcomm Snapdragon 810 with 3GB RAM), as these reflect two possible PEDs with widely varying capabilities. 
The matrix size used in tasks $t_{1}$ and $t_{2}$ are 1000$\times$1000 and 100$\times$100 for MacBook Pro and Nexus 6P, respectively. From Figure \ref{fig:pairwise}a and \ref{fig:pairwise}b, we see that the average execution latency has a linear relationship with the number of (same type) tasks running on the edge device for both platforms by looking at $T(t_1,k*t_1)$ and $T(t_2,k*t_2)$. Moreover, we also see that the lines representing $T(t_2, j*t_1+k*t_2)$ and $T(t_2,j*t_1)+T(t_2,k*t_2)$ have almost complete overlap. The same pattern is also observed for $T(t_1, j*t_1+k*t_2)$ and $T(t_1,j*t_1)+T(t_1,k*t_2)$, which verifies our assumption that the interference pattern is additive.

\Subsection{Edge device profiling}
To test \name's joint optimization of end-to-end latency and probability of failure under edge devices with various capabilities, we profiled 7 different AWS EC2 instances and a Macbook Pro for the 4 applications used in our simulation. The detailed configurations are shown in Table \ref{tbl:ec2_config}. Note that different configurations emulate different compute power, memory and number of CPUs, which result in different interference patterns for each device.

\begin{table}[h]
\caption {Edge device configuration}
\label{tbl:ec2_config}
\centering
\resizebox{\columnwidth}{!}
{%
\begin{tabular}{|l|l|l|l|l|}
\hline
\multicolumn{1}{|l|}{\text{\bf ED id}}
& \multicolumn{1}{l|}{\bf{Instance type}}
& \multicolumn{1}{l|}{\bf{(v)CPUs}}
& \multicolumn{1}{l|}{\bf{Memory(GB)}}
& \multicolumn{1}{l|}{\bf{Frequency(GHz)}}\\
\hline
ED0 & Macbook Pro 2017 & 2 & 8 & 3.1 \\  
\hline
ED1 & t2.xlarge & 4 & 16  & 2.3 \\ 
\hline
ED2 & t2.2xlarge & 8 & 32  & 2.3 \\ 
\hline
ED3 & t3.xlarge & 4 & 16  & 2.5 \\ 
\hline
ED4 & t3a.xlarge & 4 & 16  & 2.2 \\ 
\hline
ED5 & c5.2xlarge & 8 & 16 & 3.4 \\ 
\hline
ED6 & c5.4xlarge & 16 & 32  & 3.4 \\ 
\hline
ED7 & t3.2xlarge & 8 & 32  & 2.5 \\ 
\hline
\end{tabular}}
\end{table}

\begin{figure}[ht]
\includegraphics[scale=0.23]{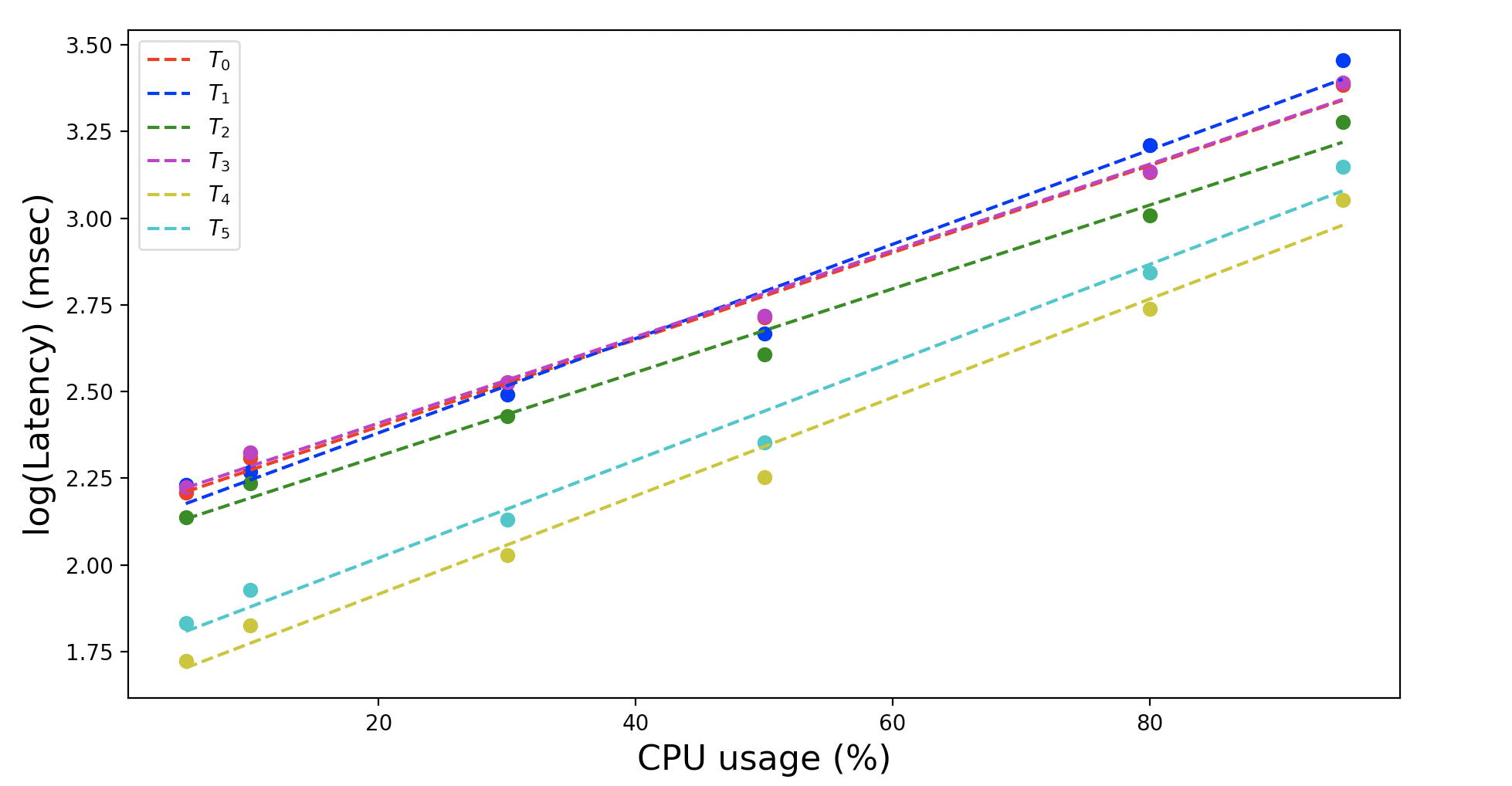}
\centering
\caption{\textbf{Profiling data on ED1 (t2.xlarge) shows the linear relationship between log(latency) and the cpu usage.}}
\label{fig:linear_cpu}
\end{figure}

Every pair of $m, c$ values (task interference parameter) between every two tasks are profiled on each of the eight edge devices. 
In addition, LaTS~\cite{lats} makes an assumption that there is a parametric model that captures the relationship between latency and CPU usage. For our baseline experiments with LaTS, we have to determine this parametric relationship in our experimental setting. Therefore, we also profiled the relationship between the CPU usage and task execution latency and fitted a linear regression model to capture such a relationship.
Figure \ref{fig:linear_cpu} shows the profiling data collected on the t2.xlarge EC2 instance at 5 different CPU usage levels for different task types. From this result, we conclude that there is a linear relationship between the log(latency) and the CPU usage. We use this linear relationship for latency estimation for our experiments with the baseline LaTS.

\begin{figure}[ht]
        \centering
        \begin{subfigure}[h]{0.37\columnwidth}  
            \centering 
            \includegraphics[width=\columnwidth]{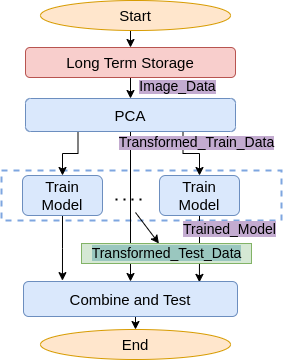}
            \caption[]{LightGBM}
            \label{fig:lightbgm}
        \end{subfigure}
        \hspace*{1.2em}
        \begin{subfigure}[h]{0.37\columnwidth}  
            \centering 
            \includegraphics[width=\columnwidth]{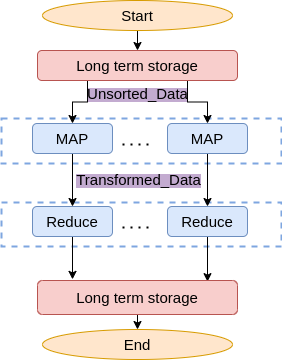}
            \caption[]{Map Reduce Sort}
            \label{fig:mrs}
        \end{subfigure}
        \hfill
        \begin{subfigure}[h]{0.37\columnwidth}
            \centering
            \includegraphics[width=\columnwidth]{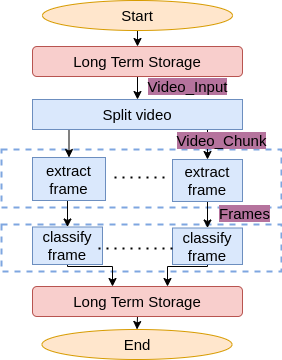}
            \caption[] {Video Analytic}
            \label{fig:video_ana}
        \end{subfigure}
        \hspace*{1.2em}
        \begin{subfigure}[h]{0.37\columnwidth}
            \centering
            \includegraphics[width=\columnwidth]{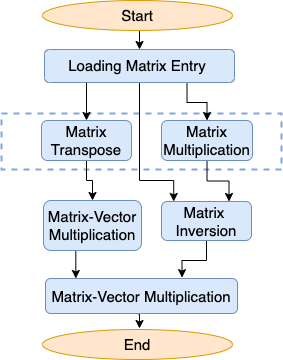}
            \caption[] {Matrix Computation}
            \label{fig:trivial}
        \end{subfigure}
        \caption[]
        {DAG applications under test} 
        \label{fig:test_app}
\end{figure}

\Subsection{Testing Applications}\label{sec:testing_app}
The simulator for our proposed framework \name is built in Python and four applications (Figure \ref{fig:test_app}) from different application domains such as  machine learning (LightGBM), data analytics (Mapreduce sort), mathematics (Matrix computation) and video analytics, that span a variety of dependency levels among tasks (DAGs) have been used for testing the generality of \name. 

We now describe our four applications and corresponding DAGs representing dependency among tasks.

\noindent\textbf{(1) LightGBM:} Figure \ref{fig:lightbgm} shows a DAG representation of the application which trains decision trees and combines them to form a random forest predictor. Such application first reads the training examples and performs dimension reduction (PCA). Then, a user-specified number of functions train the decision trees in parallel (every function randomly selects 90\% for training, 10\% for validation). In the end, all trained models are collected and combined to get tested on held-out test data. The inputs of such application are the handwritten images databases: NIST(800K images) and MNIST (60K images).

\noindent\textbf{(2) MapReduce Sort:} Figure \ref{fig:mrs} shows a DAG representation of the MapReduce sort. At the first stage, the parallel mappers (MAPs) fetch input data and generate intermediate files. In the next stage, the reducers sort the intermediate file and write the result back to the storage.

\noindent\textbf{(3) Video Analytics:} Figure \ref{fig:video_ana} shows a DAG representation of the video analytics application. At the first stage, the input video is split into multiple chunks, then a significant frame is extracted from each chunk. Eventually, the significant frame will be used for classification.

\noindent\textbf{(4) Matrix Computation:} 
Figure \ref{fig:trivial} is a matrix computation application, where all tasks are heavy matrix computations. Such setup can be mapped into various mathematics applications. In particular, the tasks used in this application are matrix inversion, matrix-matrix multiplication and matrix-vector multiplication.  

\Subsection{Baseline Systems}
We compare our orchestration scheme with the following baselines:

\noindent\textbf{LAVEA~\cite{lavea}:} LAVEA is a system built to offload computation between clients and edge nodes to provide low-latency video analytics.
We compare our scheme with their best performing scheme, Shortest Queue Length First (SQLF), which tries to balance the number of tasks running on each edge device.

\noindent\textbf{Petrel~\cite{petrel}:} Petrel is a randomized load balancing framework that utilizes the strategy of 
'the power of two choices'~\cite{power_o2}. The framework randomly selects two edge devices and offloads the task to the one that has the lowest expected service time.

\noindent\textbf{LaTS~\cite{lats}:} LaTS is a latency-aware task scheduling heuristic that distributes tasks to edge devices based on the predicted execution latency of each task through a model which characterizes the relationship between execution latency and CPU usage of the edge device node.

\noindent\textbf{Round Robin:} In this scheme, the tasks in each application instance are distributed to edge devices present in network in round-robin manner.

\noindent\textbf{Random:} In this scheme, the task will be distributed to edge devices present in network randomly.

\Subsection{Performance Metrics}

\noindent \textbf{Service Time}: We define the service time for an application instance scheduled by the orchestrator as the end-to-end latency, starting from the execution of the first task until the last task finishes its execution.
In our simulation, application instances may arrive in a clustered manner, 
which can cause tasks to accumulate on edge devices and result in longer end-to-end latency for some instances. Therefore, the average service time for a single application instance across all application instances of all applications throughout the entire simulation period is used in our measurement and is measured in seconds. 

\noindent \textbf{Probability of Failure (PF):} The probability of failure for an application instance is defined as $1-P_{(all\_success)}$, where $P_{(all\_success)}$ denotes the probability that all tasks composing the application instance (not counting the replicas) are executed successfully. Tasks can fail due to  sporadic availability of edge devices or tasks taking abnormally long to execute (e.g., a person leaves the room with his laptop during the middle of the task execution).

\Subsection{Evaluation of Heterogeneity}

Recall that heterogeneity across edge devices is captured in our work through various computing powers (Section \ref{sec: experiment}-B) of different edge devices and a mix of PEDs and CEDs. 


\begin{figure}[h]
\centering
\includegraphics[scale=0.20]{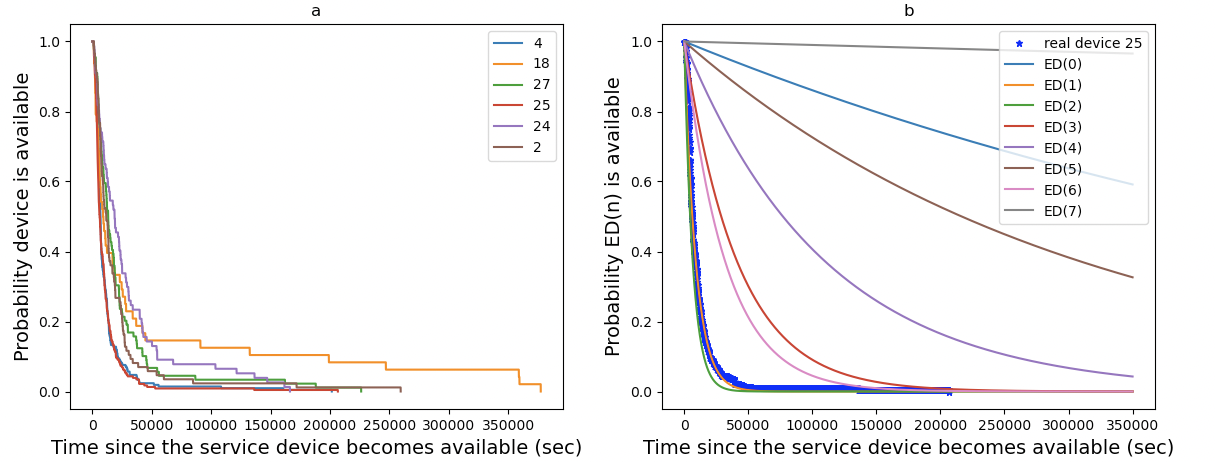}
\caption{\textbf{Availability of edge device throughout the simulation time. (a) Probability of 5 random devices' availability in mobility data collected from a university campus~\cite{mobility} 
(b) Probability of device availability used in simulations.}}
\label{fig:pf}
\end{figure}

To show the impact of heterogeneity, we used the exponential function $P(ED_{i}) = e^{-\lambda t}$ to simulate the sporadic availability 
of different edge devices with different failing rates (i.e., different $\lambda$s) for different devices. The exponential form makes sense as the average probability of failure increases as simulation time passes. 

This experiment is meant to validate that the exponential model can be used as a good prediction of the probability of availability of edge devices and the set of $\lambda$ values used in our simulation are realistic. We used the mobility trace from~\cite{mobility} to validate our assumption. The mobility data is collected over one month (Feb 7th - Mar 7th, 2018) with 50 users on a university campus. 
Each user was performing their daily routine while their smartphones were running tasks for collecting sensor data, such as geolocation. 
The missing data points in the data sets indicate students turned off their devices, have no network access or quit the data collection program for their own reasons. By analyzing the mobility data, we show the corresponding results in Figure \ref{fig:pf}a, which shows the change of probability of availability since it first becomes available. 

\begin{table}[h]
\caption {Failure rates $\lambda$ used in simulation. $\lambda_1 = $ Mix, $\lambda_2 = $ CED, $\lambda_3 = $ PED.} 
\label{tbl:hyb_related_work}
\centering
\resizebox{\columnwidth}{!}
{%
\begin{tabular}{|l|l|l|l|l|l|l|l|l|}
\hline
    \multicolumn{1}{|l|}{\text{\bf ED devices}}
& \multicolumn{1}{l|}{\bf{ED0}}
& \multicolumn{1}{l|}{\bf{ED1}}
& \multicolumn{1}{l|}{\bf{ED2}}
& \multicolumn{1}{l|}{\bf{ED3}}
& \multicolumn{1}{l|}{\bf{ED4}}
& \multicolumn{1}{l|}{\bf{ED5}}
& \multicolumn{1}{l|}{\bf{ED6}}
& \multicolumn{1}{l|}{\bf{ED7}}\\
\cline{1-8}
\hline
$\lambda_{1}$ & $1.5*10^{-6}$ & $1.1*10^{-4}$ & $1.5*10^{-4}$ & $2.4*10^{-5}$ & $9*10^{-6}$ & $3.2*10^{-6}$ & $3.1*10^{-5}$ & $1*10^{-7}$ \\  
\hline
$\lambda_{2}$ & $1.5*10^{-5}$ & $1.1*10^{-5}$ & $1.5*10^{-5}$ & $1.1*10^{-5}$ & $1.8*10^{-5}$ & $1.2*10^{-5}$ & $1.0*10^{-5}$ & $2.0*10^{-5}$ \\  
\hline
$\lambda_{3}$ & $1.5*10^{-4}$ & $1.1*10^{-4}$ & $1.5*10^{-4}$ & $2.4*10^{-4}$ & $9*10^{-4}$ & $3.2*10^{-5}$ & $1.0*10^{-4}$ & $9.0*10^{-4}$ \\  
\hline
\end{tabular}}
\end{table}

Table~\ref{tbl:hyb_related_work} shows different sets of $\lambda$ values that have been used in our simulations. $\lambda_{1}$ simulates the scenario of a mixture of PEDs and CEDs. $\lambda_{2}$ represents the scenario when there are only CEDs present and $\lambda_{3}$ represents when there are only PEDs available. We specifically plot the set of $\lambda_{1}$ (mix of CEDS and PEDs) and the real-world data collected from the real-world participants 
in Figure \ref{fig:pf}b, and we see that the model for $ED_{6}$ fits the real-world data well which verifies our assumption that exponential function can be used as a good prediction for device availability with careful selection of failure rates $\lambda$.

\begin{figure}[ht]
        \centering
        \begin{subfigure}[h]{0.5\columnwidth}  
            \centering 
            \includegraphics[width=\columnwidth]{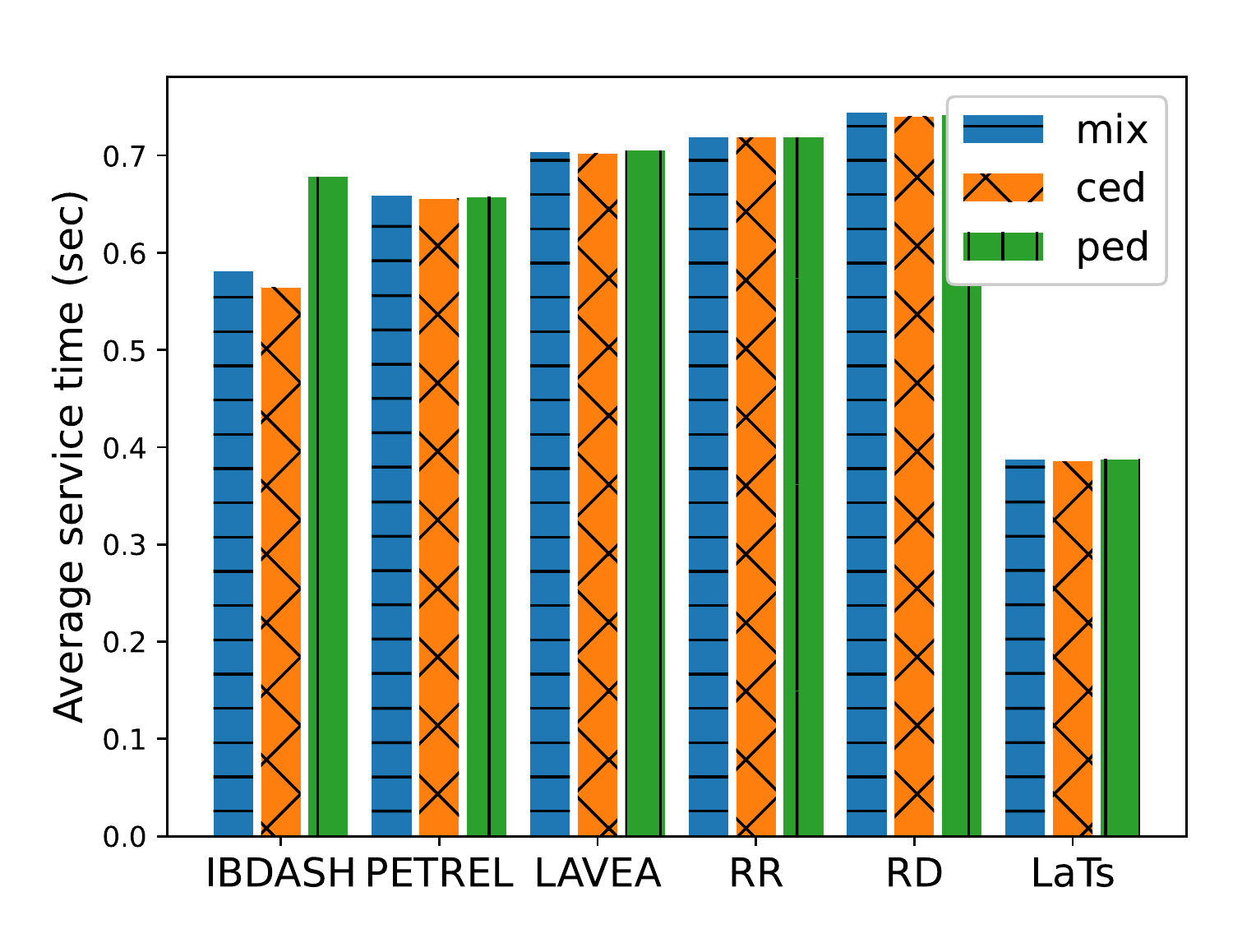}
            \caption[]{Video analytics}
            \label{fig:va_servicetime}
        \end{subfigure}
        \hspace*{-1.3em}
        \begin{subfigure}[h]{0.5\columnwidth}  
            \centering 
            \includegraphics[width=\columnwidth]{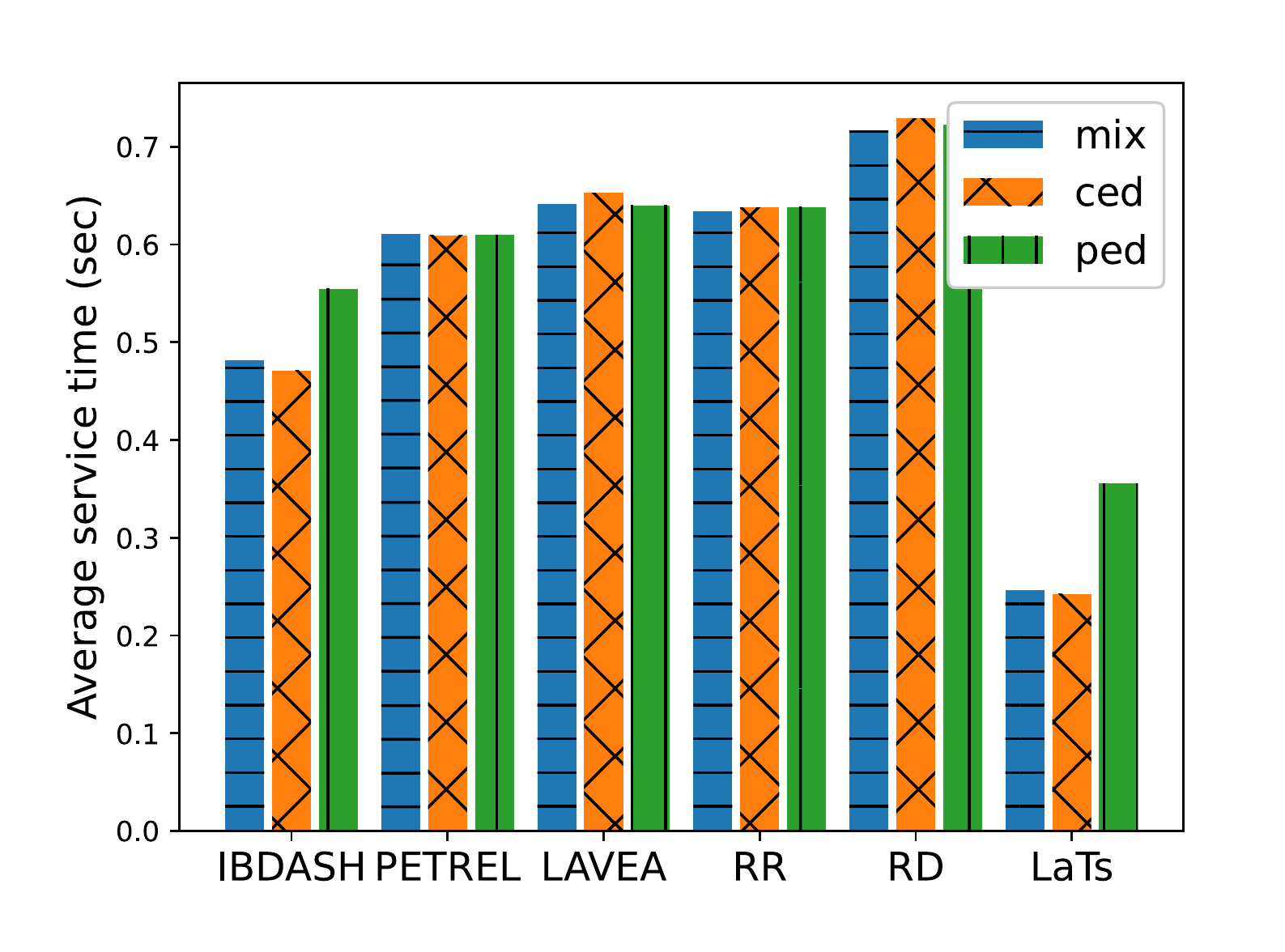}
            \caption[]{LightGBM}
            \label{fig:lightbgm_servicetime}
        \end{subfigure}
        \begin{subfigure}[h]{0.5\columnwidth}
            \centering
            \includegraphics[width=\columnwidth]{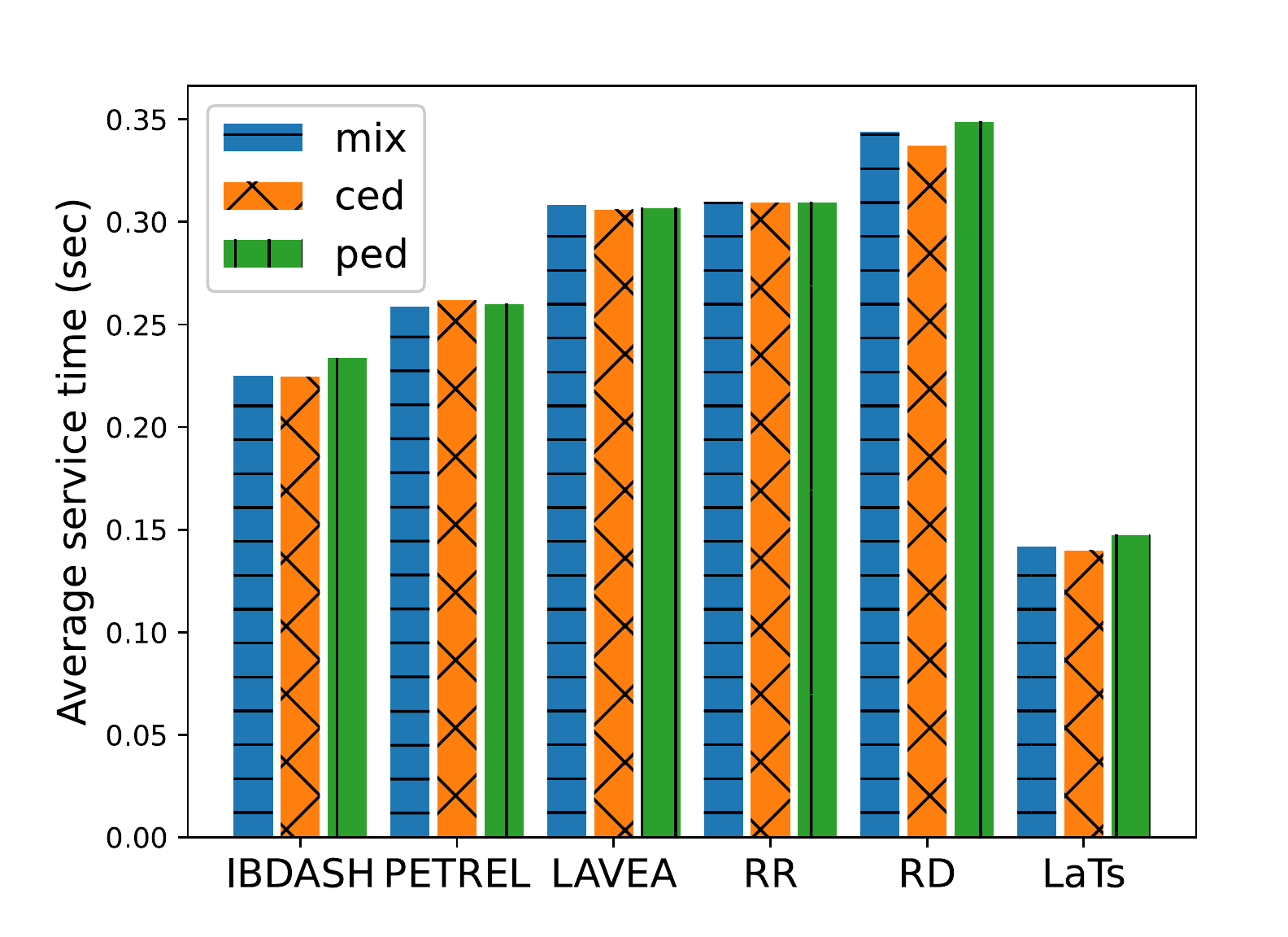}
            \caption[] {Map reduce sort}
            \label{fig:mr_servicetime}
        \end{subfigure}
        \hspace*{-1.3em}
        \begin{subfigure}[h]{0.5\columnwidth}
            \centering
            \includegraphics[width=\columnwidth]{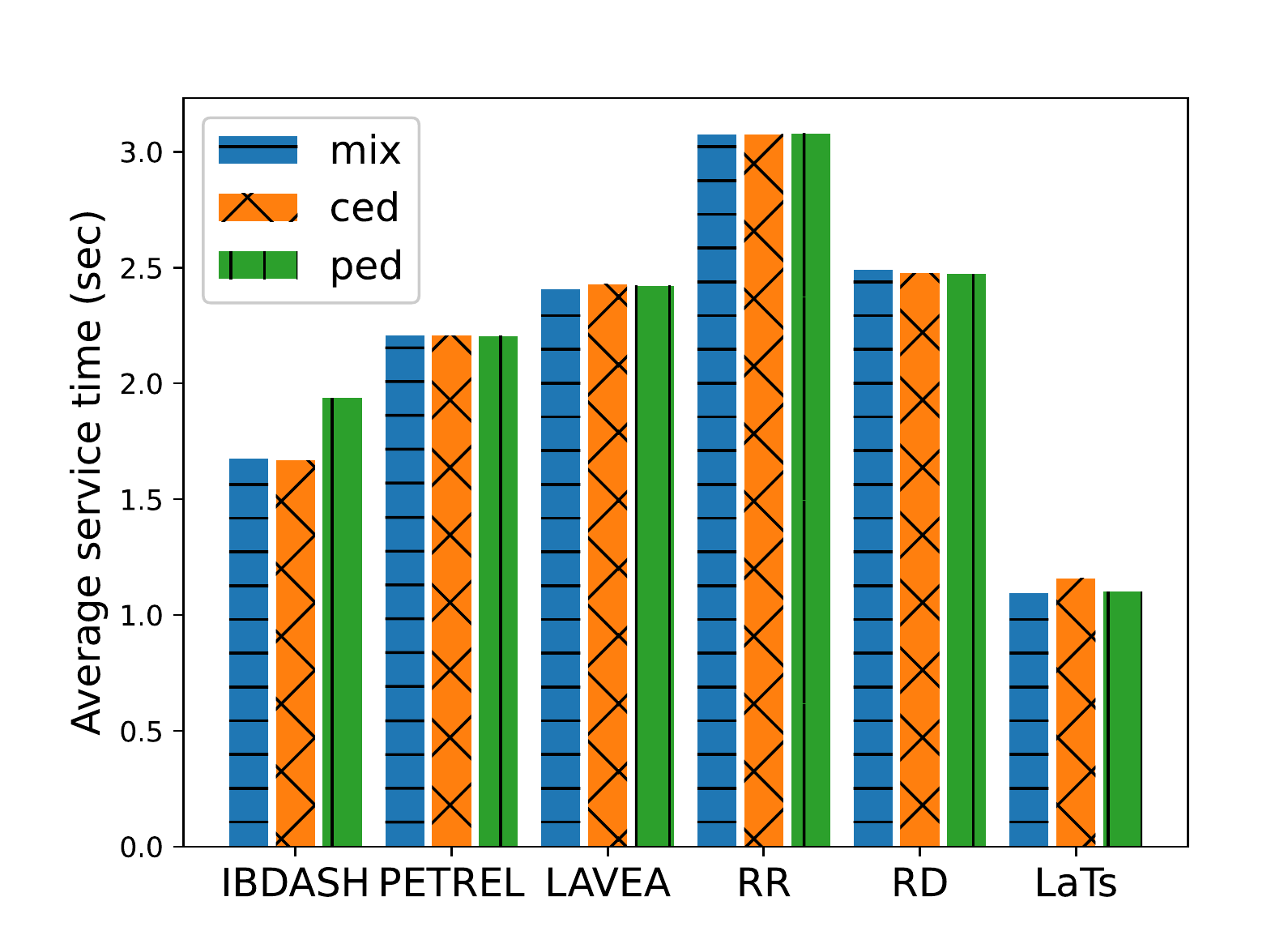}
            \caption[] {Matrix computation}
            \label{fig:trivial_servicetime}
        \end{subfigure}
        \caption[]
        {\textbf{The average service time for all 4 testing applications under 6 different orchestration schemes. \name outperforms other schemes under all tests (except LaTS)}} 
        \label{fig:service_time_four}
    \end{figure}

\Subsection{Evaluation of End-to-End Latency}
In our experiment, we repeated a 15s simulation cycle 20 times giving a total simulation time of 5 minutes. In each cycle, 1000 application instances arrive randomly clustered within the initial 1.5s and there are 100 edge devices uniformly distributed among the 8 device classes listed in Table \ref{tbl:ec2_config}. Their corresponding interference coefficients are collected from real-world experiments. It can be observed from Figure \ref{fig:service_time_four} that the average service time of \name outperforms other orchestration schemes except for LaTS under all three scenarios (CEDs, PEDs, the mix of CEDs and PEDs (50\%:50\%)) for all applications due to its awareness of the co-located task interference. The reason that LaTS outperforms \name in execution latency comparison is that LaTS allocates the majority of tasks to a single powerful device. However, if that device were to become unavailable, then the performance of LaTS will suffer drastically. 

\Subsection{Evaluation of Probability of Failure}


\begin{figure}[ht]
        \centering
        \begin{subfigure}[h]{0.5\columnwidth}  
            \centering 
            \includegraphics[width=\columnwidth]{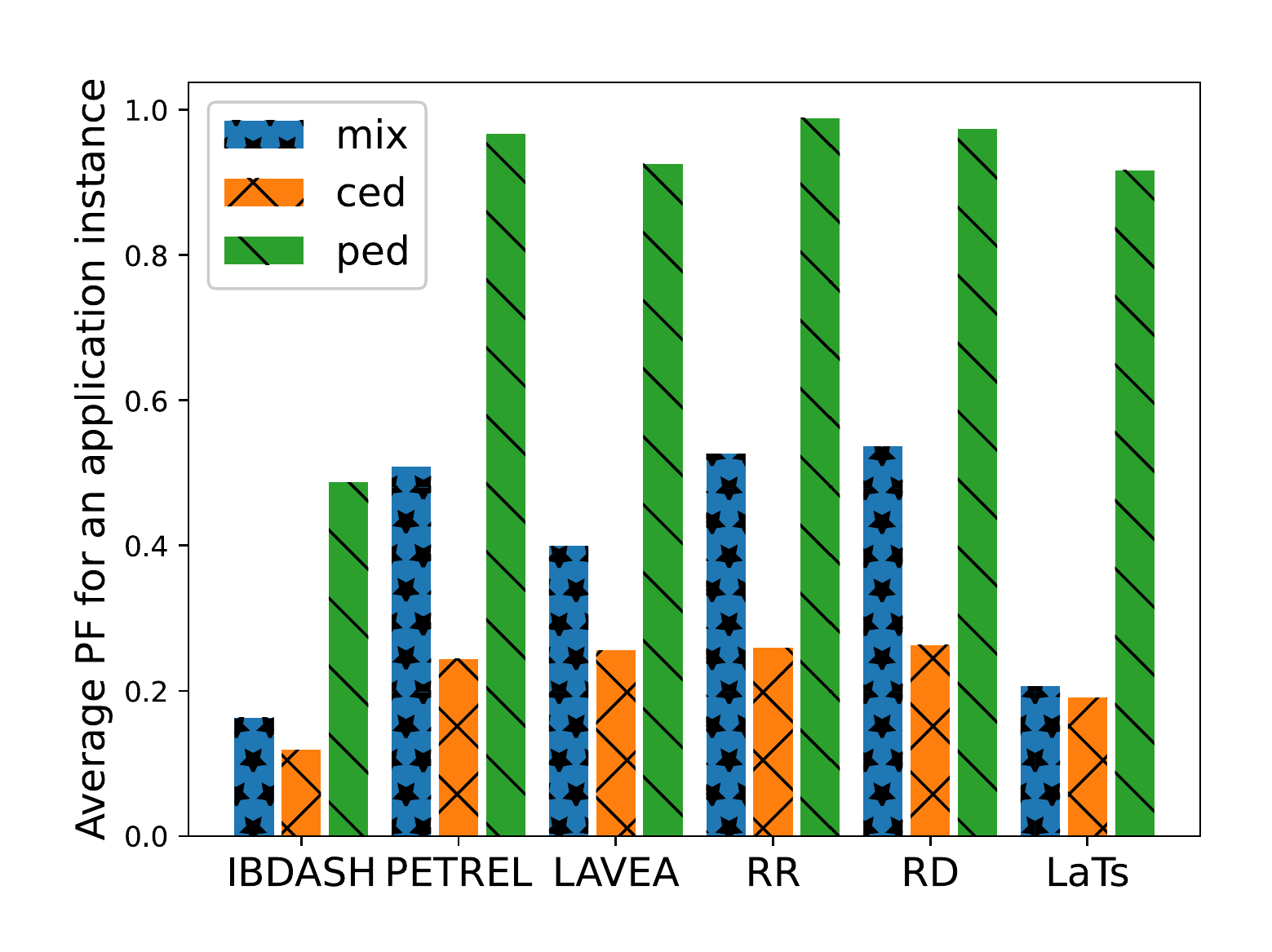}
            \caption[]{Video analytics}
            \label{fig:va_pf}
        \end{subfigure}
        \hspace*{-1.3em}
        \begin{subfigure}[h]{0.5\columnwidth}  
            \centering 
            \includegraphics[width=\columnwidth]{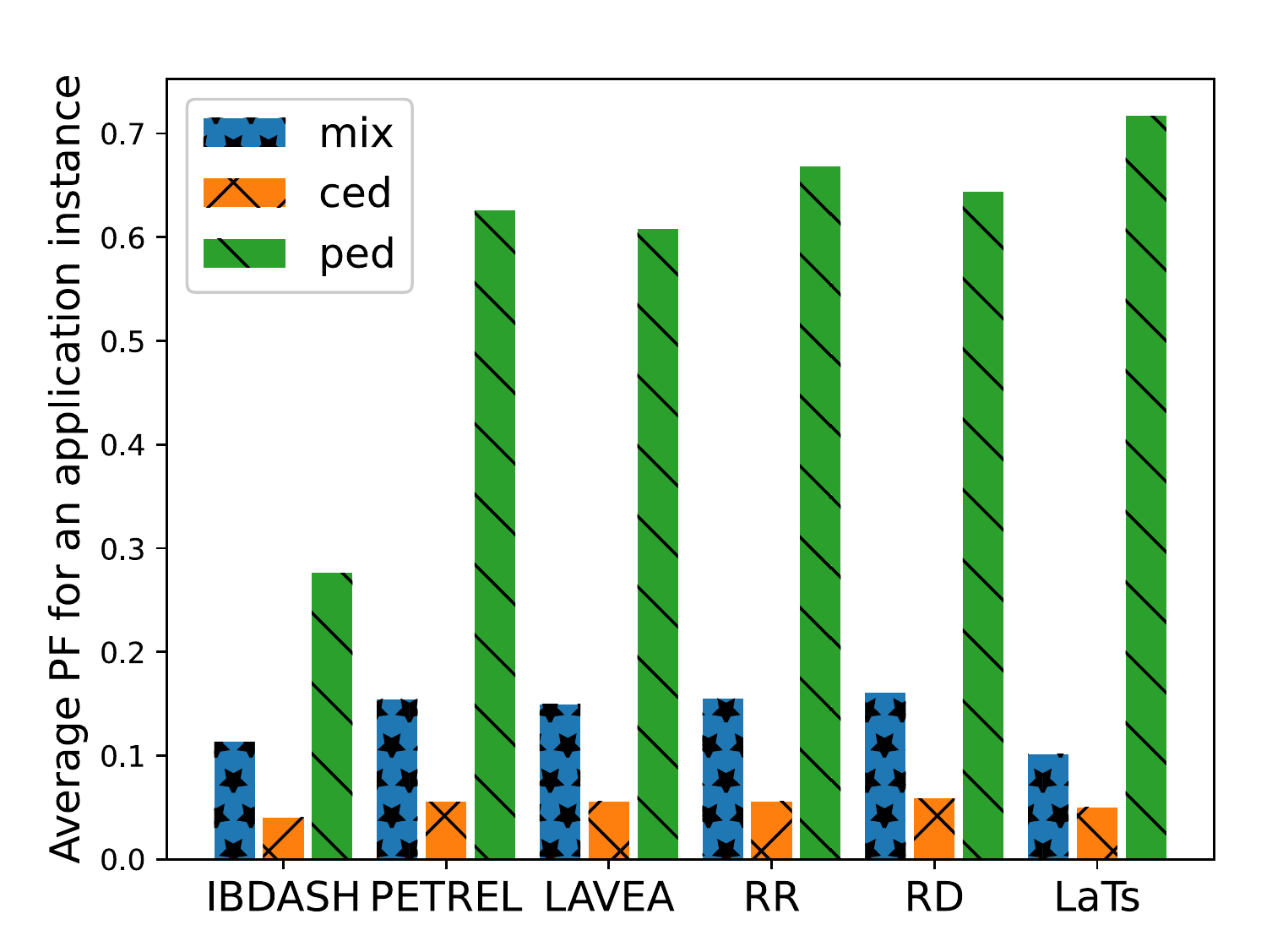}
            \caption[]{LightGBM}
            \label{fig:lightbgm_pf}
        \end{subfigure}
        \begin{subfigure}[h]{0.5\columnwidth}
            \centering
            \includegraphics[width=\columnwidth]{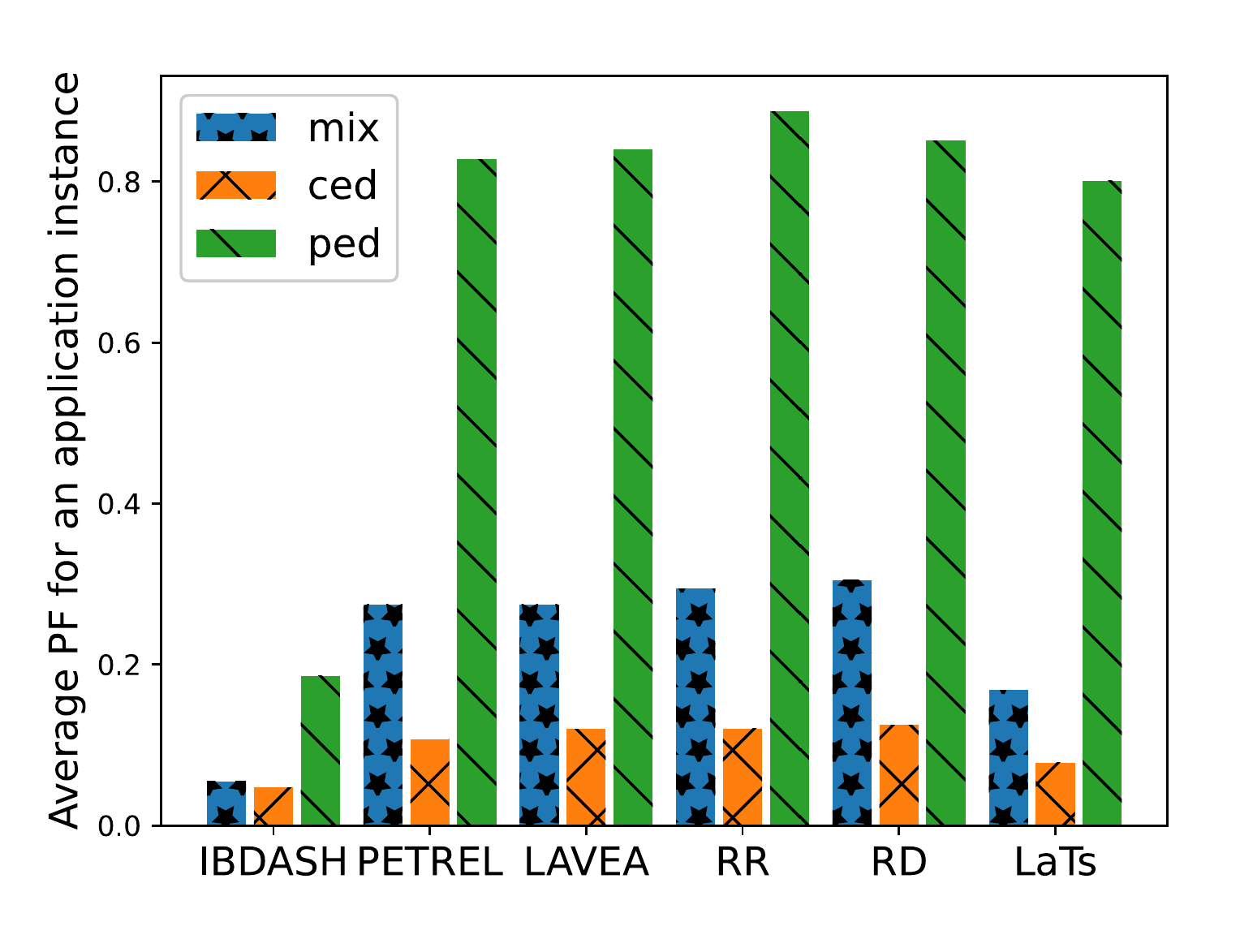}
            \caption[] {Map reduce sort}
            \label{fig:mr_pf}
        \end{subfigure}
        \hspace*{-1.3em}
        \begin{subfigure}[h]{0.5\columnwidth}
            \centering
            \includegraphics[width=\columnwidth]{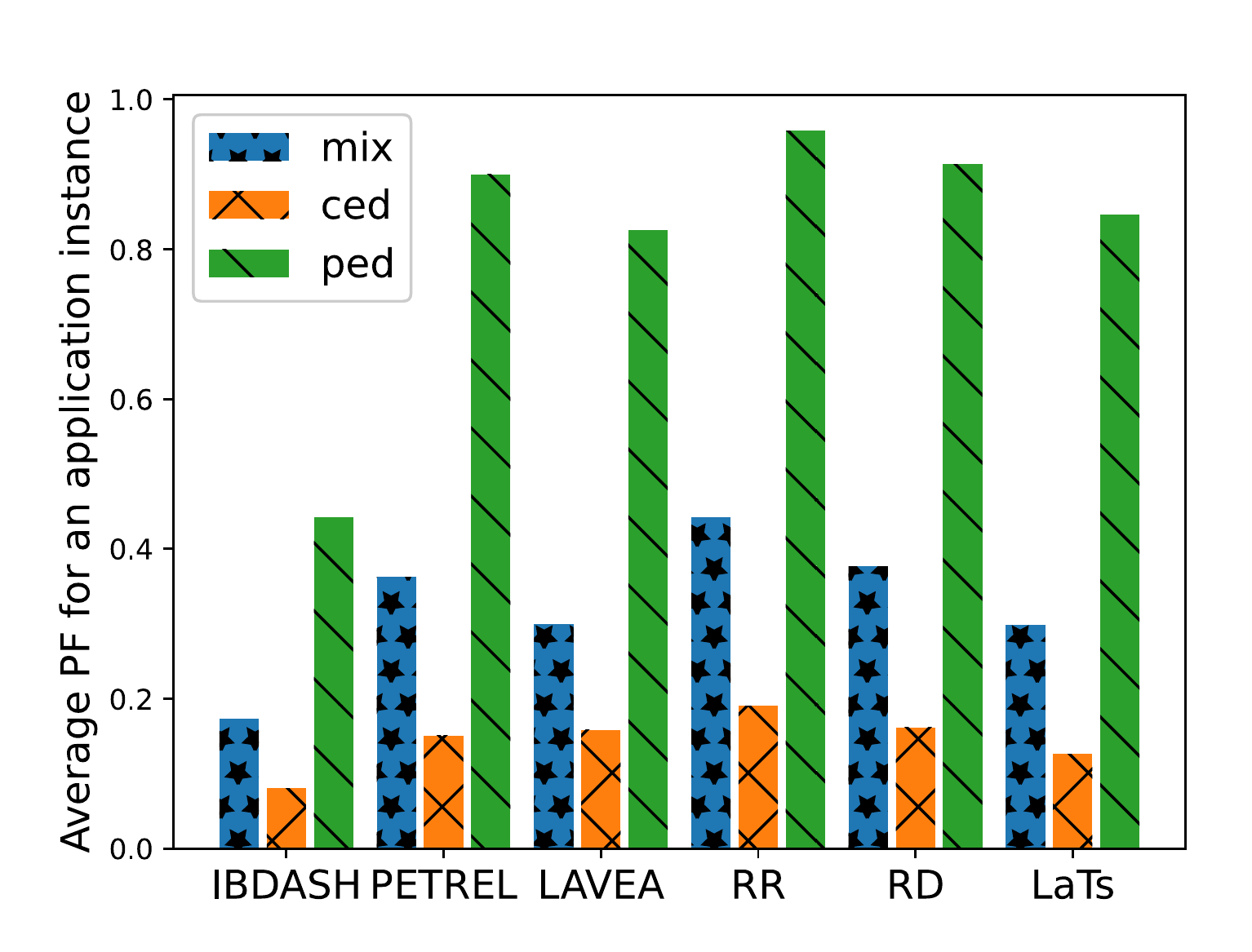}
            \caption[] {Matrix computation}
            \label{fig:trivial_pf}
        \end{subfigure}
        \caption[]
        {\textbf{Average probability of failure for 4 testing applications under 6 different orchestration schemes. \name outperforms other schemes especially when edge devices are PEDs or PEDs and CEDs combination.}}
        \label{fig:pf_of_four}
\end{figure}
    
Figure \ref{fig:pf_of_four} demonstrates the average probability of failure of application instances under six different orchestration schemes for three different scenarios. 
We see that \name outperforms other baselines under all three scenarios, especially in the scenario where edge devices are a combination of PEDS and CEDs or all are PEDs as \name offers the redundant replication to reduce the probability of failure. \name is better than LaTS by 29.7\% for mix, 58.5\% for PEDs, and 34\% CEDs on average across four applications. We emphasize that LaTS outperforms \name in rare cases since the majority of tasks are allocated to a single device. If that single device being allocated had a low probability of failure, the overall probability of failure value is low. However, as discussed earlier, this can lead to a catastrophic failure of all application instances. 

\begin{figure}[h]
\includegraphics[scale=0.29]{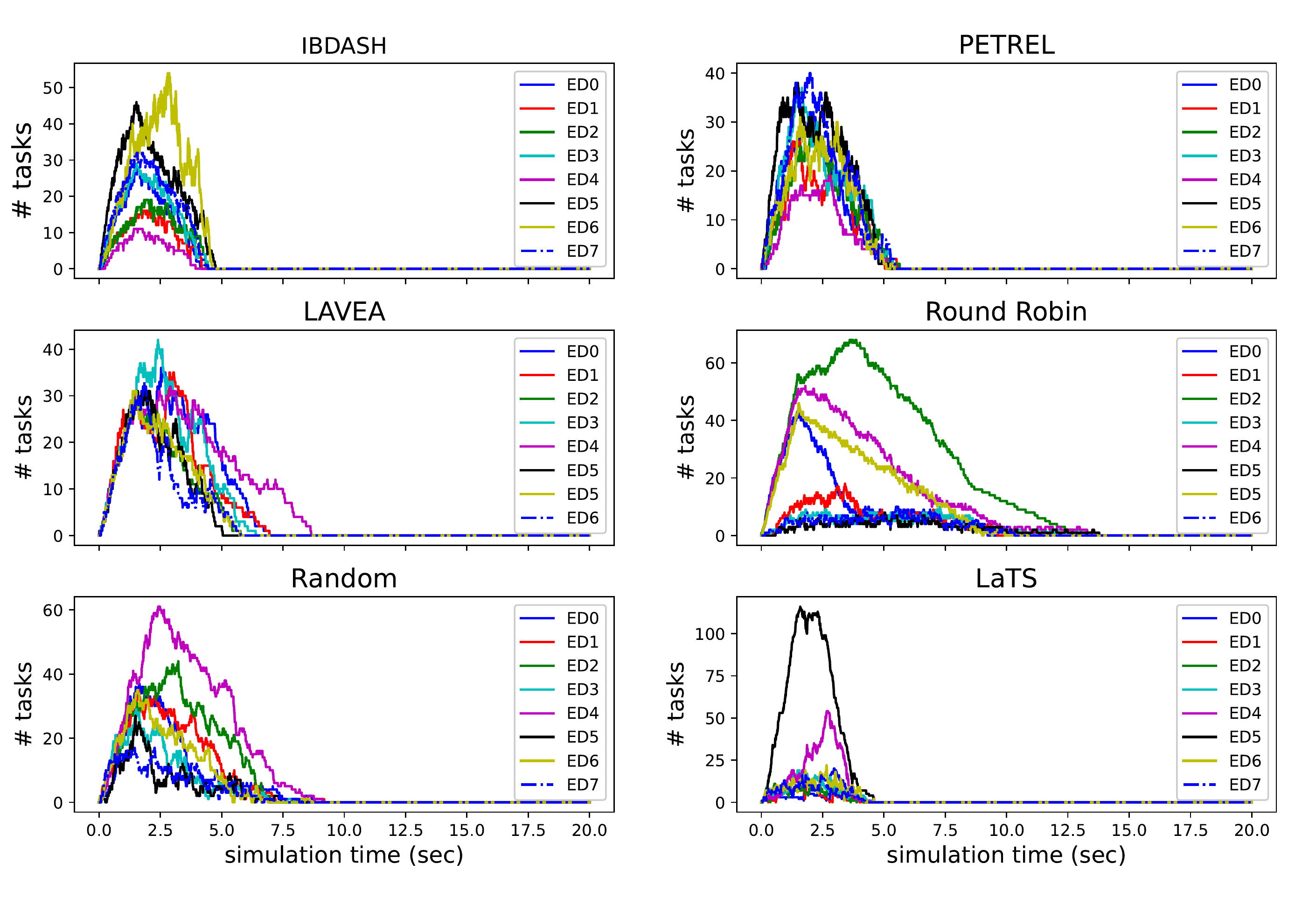}
\centering
\caption{\textbf{Plot of load on each edge device for different orchestration schemes under mixed scenario ($\lambda_1$ in Table \ref{tbl:hyb_related_work}). \name provides an even load.}}
\label{fig:load}
\end{figure}

\begin{figure}[h]
    \begin{subfigure}[h]{1\columnwidth}  
    \centering 
    \includegraphics[width=\columnwidth]{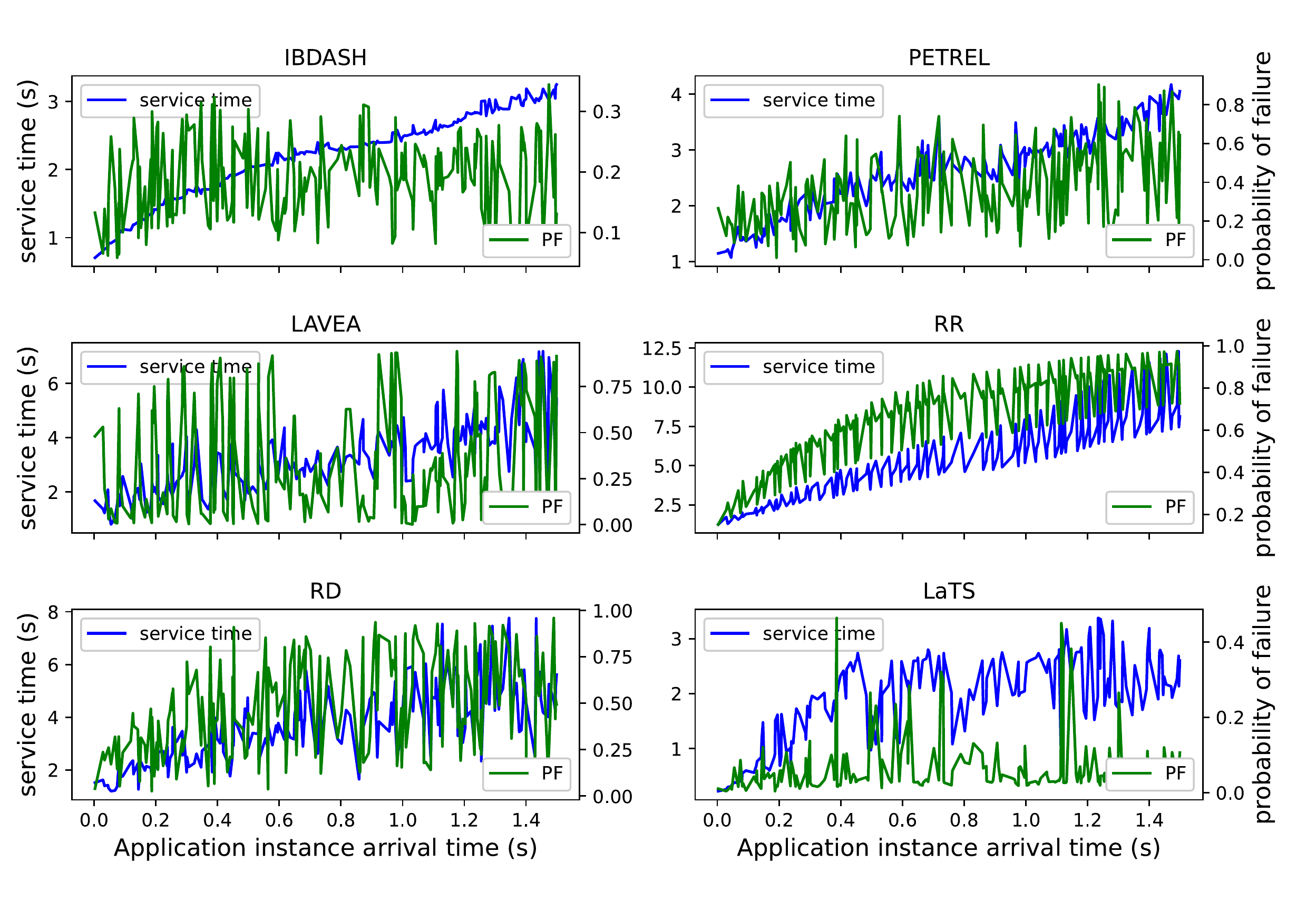}
    \end{subfigure}
    \caption{\textbf{Plot of service time and probability of failure of each individual instance for 200 application instances that arrive randomly in 1.5 second under mixed PED:CED scenario ($\lambda_1$ in Table \ref{tbl:hyb_related_work}) for six different orchestration schemes.}}
\label{fig:service_time_detail}
\end{figure}

\Subsection{Microscopic View}
To show the advantage of using \name in detail, we performed separate experiments (shown in Figure~\ref{fig:load} and Figure~\ref{fig:service_time_detail}) with 8 edge devices (one from each class) so that we can plot the loads, which is the number of tasks
on each device and examine the load distribution. 
We zoom into one of the 15s simulation cycles for this experiment.  We see that \name tends to allocate more tasks on edge devices with low interference coefficients to reduce the overall service time, in this case, devices ED5 and ED6. 
On the other hand, LAVEA~\cite{lavea} chooses to allocate the task to the edge device with the least number of running tasks, which results in the load being fairly balanced on each edge device. Petrel~\cite{petrel} chooses two edge devices randomly and allocates the task to the one which has a lower expected service time. It results in a fairly balanced load distribution as well except for those edge devices with significantly larger interference coefficients compared to others. LaTS~\cite{lats} allocates the majority of the tasks to ED5 due to its significant superiority of performance compared to others. Even though it does produce low execution latency, it results in a highly imbalanced allocation, which has negative consequences. 
Both Round Robin and Random allocation result in task accumulation on edge devices with high interference coefficients due to their fixed task distribution scheme, which leads to a long service time.

From Figure \ref{fig:service_time_detail}, we see that when the probability of failure of edge devices increases toward the end of the simulation, our orchestrator \name starts to replicate the tasks to reduce the probability of failure, which results in increasing the overall number of tasks on some edge devices (Figure \ref{fig:load}) and correspondingly the average service time for the application instance goes up due to the redundant tasks. For Petrel~\cite{petrel}, LAVEA~\cite{lavea}, the average probability of failure is higher as they do not have the extra redundancy to reduce the probability of failure. As for LaTS~\cite{lats}, it assigned most tasks to a single device, which happens to have a low probability of failure, so it shows a fairly low probability of failure.

\begin{figure}[h]
        \begin{subfigure}[h]{0.48\columnwidth}  
            \centering 
            \includegraphics[width=\columnwidth]{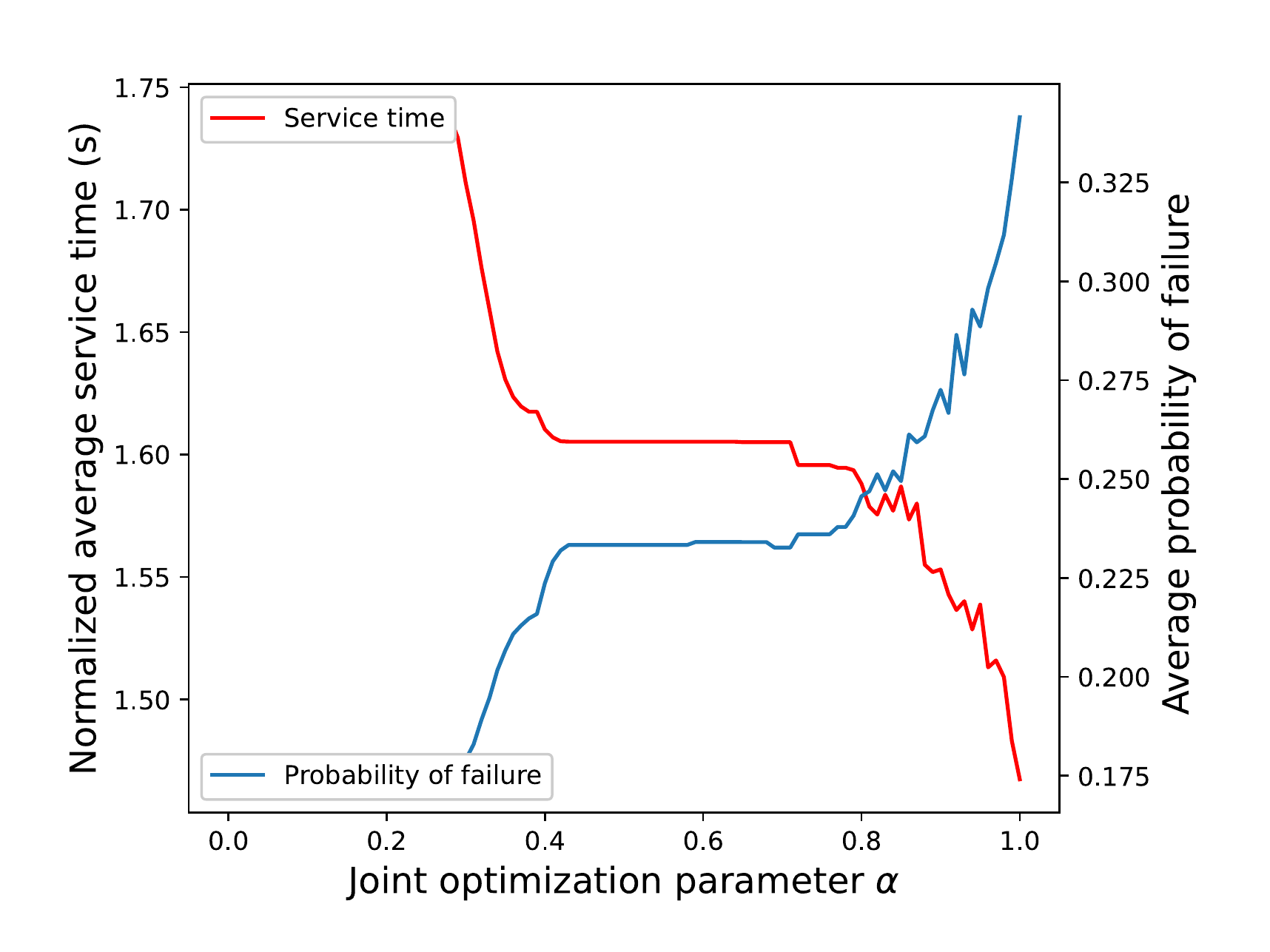}
            \caption[]{ Optimization parameter $\alpha$ sweep}
            \label{fig:sweep}
        \end{subfigure}
        \hspace*{-0.5em}
        \begin{subfigure}[h]{0.48\columnwidth}  
            \centering 
            \includegraphics[width=\columnwidth]{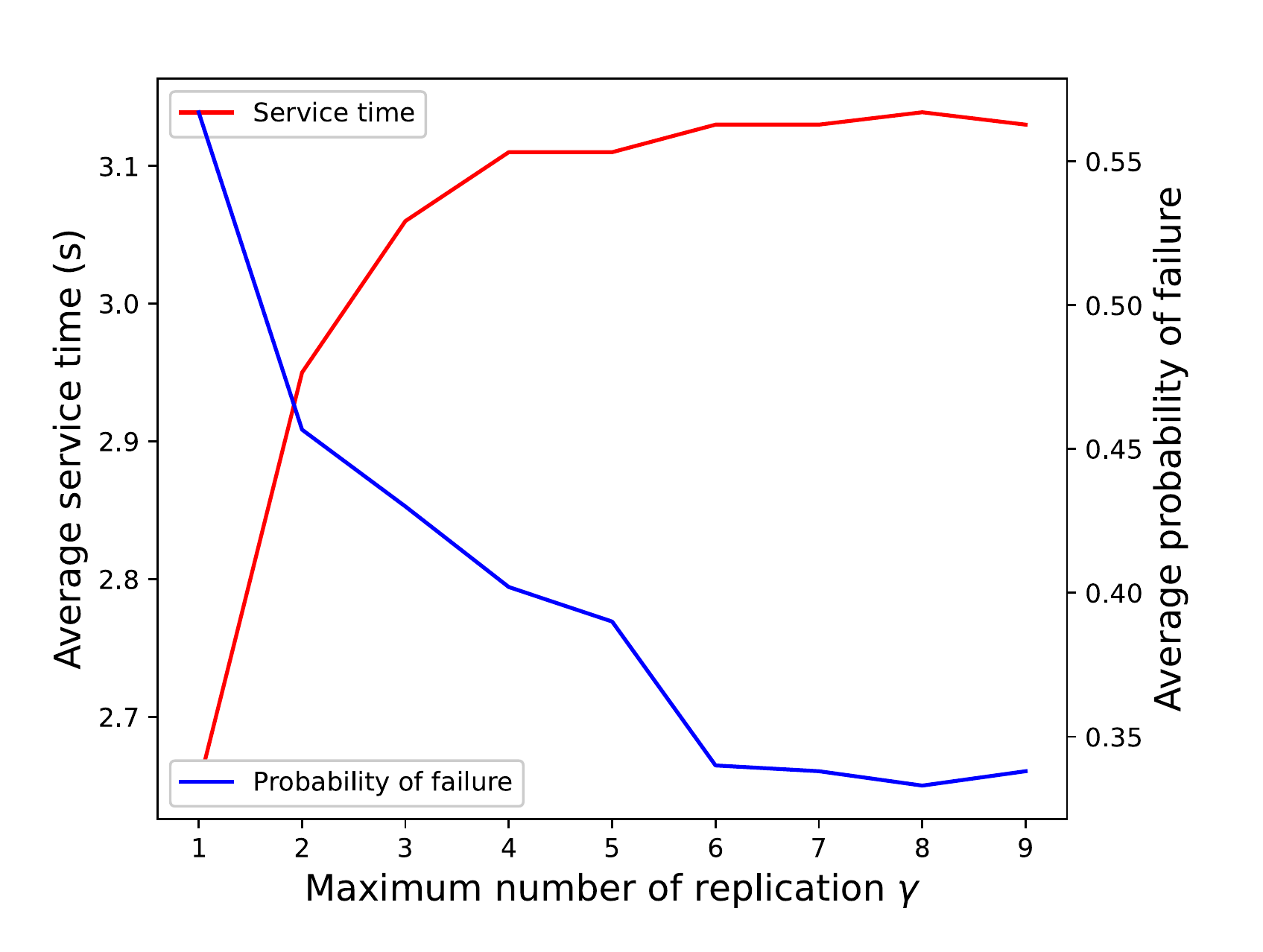}
            \caption[]{Maximum replication degree $\gamma$}
            \label{fig:repsweep}
        \end{subfigure}
        \caption[]
        {\textbf{a. Sweep of joint optimization parameter $\alpha$ from 0 to 1 in step of 0.01 @ $\beta=0.1,\gamma=3$ under $\lambda_{1}$ Table~\ref{tbl:hyb_related_work}. b. Sweep of replication degree $\gamma$ @ $\beta=0.1, \alpha=0.5$ under $\lambda_{3}$ Table~\ref{tbl:hyb_related_work}}}
        \label{fig:twosweep}
\vspace{-3mm}
\end{figure}

\Subsection{Evaluation of Joint Optimization of End-to-end Latency and Probability of Failure}
To evaluate the joint optimization of our orchestration scheme, we performed a sweep of the replication threshold $\gamma$  and (separately) of the joint optimization parameter $\alpha$.
The result is shown in Figure \ref{fig:twosweep}.

In Figure \ref{fig:sweep}, initially, when the probability of failure is assigned much more weight than the service time, the algorithm tends to optimize the probability of failure as much as it can until it meets the probability of failure threshold $\beta$ or replication degree $\gamma$. 
At around $\alpha=0.3$, \name starts to prioritize optimizing service time as more weight is given to it and shorter service time gives a better joint optimization score. Throughout the sweep, we see the general trend is that as the normalized service time decreases, the average probability of failure increases. However, there are some fluctuations in the sweep. For example, at $\alpha \approx 0.8$, we see that there is a temporary drop in the probability of failure and 
an increase in the average service time. The reason for this is that as the $\alpha$ value changes in each sweep, the task allocation changes as well. Therefore, this change in task allocation can result in fluctuations for both probability of failure and service time, but the general trend is not affected as shown in Figure \ref{fig:sweep}.

From Figure \ref{fig:repsweep}, we see that as we increase the replication degree, the average service time increases while the average probability of failure decreases, then it stays fairly stable after around 6 replications as \name is able to determine that further replications will not result in better joint optimization, therefore it stops replicating. 


%% file: Relatedwork.tex
\Section{Related Work} \label{sec: rw}

\noindent\textbf{Scheduling on the edge}: One of the most important goals in edge computing is reducing the end-to-end latency to enable latency-sensitive applications for users. Several prior works such as LaTS~\cite{lats}, LAVEA~\cite{lavea}, and Petrel~\cite{petrel} propose scheduling strategies that aim at minimizing the service time in a multi-edge collaborative environment. We have shown that \name outperforms these schemes in terms of average service time and probability of failure in a heterogeneous edge computing setting. Other works consider joint optimization of low latency and other goals under some constraints~\cite{deep_decision, data_aware}. In particular, ~\cite{deep_decision} 
focuses on the latency and accuracy optimization for video analytics under battery, network, and cost constraints.
There is a growing body of work on low-latency scheduling on the edge~\cite{zhang2019edgebatch, he2018s} and a subset considers DAG-based applications~\cite{liao2021dependency}. 
However, none of these works considers the interference among co-located tasks or intermittent availability of a subset of devices.

\noindent\textbf{Interference-based scheduling}: A few efforts study the availability of heterogeneous edge devices and interference among tasks on the same edge device~\cite{lats,tracon,score_base}. LaTS~\cite{lats} proposed to use a Latency-CPU usage model to address the interference among co-located tasks. It constantly monitors the CPU usage on each edge device and combines with the Latency-CPU usage model to schedule tasks to get the minimum predicted latency. Moreover, ~\cite{score_base} proposed a score-based edge service scheduling algorithm that evaluates network
, compute, and reliability capabilities of edge nodes
, but the drawback of such algorithm is that it requires sharing of monitoring information across all devices which is infeasible in edge computing. Compared to those frameworks, \name 
considers the heterogeneity in edge computing and requires much less information sharing among all edge devices.

\noindent\textbf{Edge device reliability and failure prevention}: There is a significant amount of work on investigating the reliability of edge devices and failure prevention~\cite{ullc,lired, relia_lat}. 
~\cite{relia_lat} conducted a small scale experiment to show the trade-offs between reliability and latency for edge nodes and server-less computing functions.~\cite{ullc} proposed three different algorithms for offloading that are based on heuristic search to reduce
the failure probability and latency, but it fails to address the interference of task co-location on the same edge device and the heterogeneity of edge devices.~\cite{lired} proposed a fault detection model based on the long short-term memory (LSTM) recurrent neural networks that are used in industrial robot manipulators. None of those frameworks addresses how to predict edge device availability. On the other hand, our work offers a model for edge device availability prediction and consequently guides the extra redundancy needed for application success.

%% file: Discussion.tex
\Section{Discussion and Future Work}\label{sec:discussion}
In this section, we present extensions of \name that would need to be implemented to handle some use cases. {\em First}, the current algorithm checks each incoming task against {\em all} available edge devices. This procedure can result in high orchestration overhead for simple tasks when many edge devices available. This problem can be addressed through edge device clustering based on (static) capabilities and (dynamic) load on each device. Then the computation overhead reduces from number of devices to number of clusters. Any of several existing techniques for edge device clustering can be used, such as~\cite{cluster1}.
{\em Second}, we hope that the execution latencies of tasks within the same stage are fairly balanced. A long latency task in a stage can delay the execution of later stage tasks. The task execution latency balancing can be achieved through further task partitioning. {\em Third}, the linearity in the task interference plots may not hold if the number of tasks running on an edge device is large enough to cause a discontinuous change such as cache spillage. In that case, a higher-order characterization (say, quadratic or piece-wise linear) of the interference plots is needed to accurately predict the execution latency. Finally, we use exponential functions to predict the sporadic availability of edge devices. Even though we validated this assumption using real-world mobility data, this may not hold in certain scenarios (say a student class schedule changes from one module to the next). This can be improved by using the history of availability of each edge device and semi-Markov process to predict availability.

%% file: Conclusion.tex
\Section{Conclusion and Takeaways}\label{sec: conclusion}
In this paper, we proposed a novel orchestration framework, \name, that enables multi-stage applications to be executed on edge computing systems. Crucially \name can incorporate personal edge devices (PEDs) along with commercial edge devices (CEDs) in executing the tasks that constitute the application. To support this, we make three novel contributions. First, \name determines the dependency among different tasks within an application represented using a DAG. Second, \name leverages PEDs while accounting for the possibility of resource contention and low and unpredictable availability of such devices. Third, \name jointly minimizes the average application execution latency (via dynamic scheduling) and application failure likelihood (via task replication). We evaluated \name with four applications that span various DAG structures, with unit measurements of real application tasks on real devices. We compared \name with three  state-of-the-art edge scheduling solutions, LAVEA, Petrel, and LaTS. We observe that \name yields an average reduction of 14\% on the service time of applications and reduces the  average  probability  of  failure  for  the  applications  by  41\%. 

There are three takeaways from our work that are of general applicability to edge computing systems. {\em First}, it is possible to leverage highly heterogeneous devices to compose a usable, i.e., low-latency and reliable, edge computing platform. 
{\em Second}, it is feasible to use unmanaged edge devices (called PEDs here) to create the usable edge computing platform, {\em if} these are combined with commercially managed devices. 
{\em Third}, a usable edge computing platform, unlike a cloud computing platform, must manage anticipated failures by proactively replicating tasks as these are far more likely than in the cloud computing world.